\documentclass[12pt,a4paper]{article}
\usepackage[margin=2cm]{geometry}
\usepackage{comment}
\usepackage{amsmath,amssymb,extarrows,graphicx,subfigure,setspace}
\usepackage{cite}
\usepackage{slashed}
\usepackage{color}
\usepackage{braket}
\usepackage{float}
\makeatother
\usepackage{pdfpages}
\usepackage{blindtext}
\usepackage{amsfonts}
\usepackage{tensor}
\usepackage{graphicx}
\usepackage{pict2e}
\usepackage{amssymb, amsmath,environ}
\usepackage[
colorlinks=true,
linkcolor=blue,
urlcolor=blue,
filecolor=blue,
citecolor=red,
pdfstartview=FitV,
pdftitle={},
pdfauthor={},
pdfsubject={},
pdfkeywords={},
pdfpagemode=None,
bookmarksopen=true
]{hyperref}

\usepackage{epsfig}

\usepackage{hyperref}

\newcommand{\be}{\begin{equation}}
	\newcommand{\bea}{\begin{eqnarray}}
		\newcommand{\eea}{\end{eqnarray}}
	\newcommand{\ba}{\begin{array}}
		\newcommand{\ea}{\end{array}}
	\newcommand{\ee}{\end{equation}}
\newcommand{\bes}{\begin{equation*}}
	\newcommand{\beas}{\begin{eqnarray*}}
		\newcommand{\eeas}{\end{eqnarray*}}
	\newcommand{\bas}{\begin{array*}}
		\newcommand{\eas}{\end{array*}}
	\newcommand{\ees}{\end{equation*}}

\numberwithin{equation}{section}
\begin{document}
	\onehalfspacing
	\noindent
	
	\begin{titlepage}
		\vspace{10mm}
		
		\vspace*{20mm}
		\begin{center}
			{\Large {\bf Symmetry-Resolved Entanglement Entropy for Local and Non-local QFTs }\\
			}
			
			\vspace*{15mm}
			\vspace*{1mm}
			{ Reza Pirmoradian $^{a,b,c}$ and M Reza Tanhayi $^{a,d}$}
			
			\vspace*{.3cm}
			
			{\it 	${}^a $School of Physics, 	${}^{b} $School of Particles and Accelerators,   Institute for Research in Fundamental Sciences (IPM), 
					P.O. Box 19568-36681, Tehran, Iran
								
				${}^c $ Ershad Damavand, Institute of Higher Education (EDI)\\
				P.O. Box 14168-34311, Tehran, Iran}\\

			{\it 
			${}^d $ 	Department of Physics, Central Tehran Branch, Islamic Azad University  (IAU)\\ P.O. Box 14676-86831,
				Tehran, Iran}\\
			
			\vspace*{0.5cm}
			{E-mails: {\tt rezapirmoradian@ipm.ir, mtanhayi@ipm.ir}}
			
			\vspace*{1cm}
		\end{center}

		\date{\today}
		
		\begin{abstract} 
		In this paper, we investigate symmetry-resolved entanglement entropy (SREE) in free bosonic quantum many-body systems. Utilizing a lattice regularization scheme, we compute symmetry-resolved R\'enyi entropies for free complex scalar fields and a specific class of non-local field theories, where entanglement entropy (EE) exhibits volume-law scaling. We present effective and approximate eigenvalues for the correlation matrix used in computing SREE and demonstrate their consistency with numerical results. Furthermore, we explore the equipartition of EE, verifying its effective behavior in the massless limit. Finally, we comment on EE in non-local quantum field theories and provide an explicit expression for the symmetry-resolved R\'enyi entropies.

			
			
		\end{abstract}
		
	\end{titlepage}

	\tableofcontents

	\newpage
	
	\section{Introduction} \label{intro}
	
	Entanglement, which describes non-local correlations between quantum systems, is one of the most fundamental topics in physics, and its concept emerged during the evolution of quantum mechanics \cite{Horodecki:2009zz}. Entanglement plays an important role in the study of various branches from theoretical to experimental physics, e.g., quantum information theory \cite{chuang,upper,Os02b}, condensed matter physics\cite{Vidal:2002rm,Calabrese:2004eu,Kitaev:2005dm,Casini:2006es}, gravity especially black holes in high energy physics \cite{bombelli,cw,srednicki} and the holographic principle\cite{ holo,Takayanagi:2012kg,Faulkner:2013ana,Headrick:2019eth,Solodukhin:2006xv, Pirmoradian:2021wvo, Tanhayi:2017wcd,Tanhayi:2016uui}.  In this way, measuring entanglement remains a challenge and so far, various quantities have been introduced to quantify entanglement measurements, the most well-known of which for pure quantum states is the EE \cite{Bennett:1995tk}.\\
	In an arbitrary quantum system, EE can be considered as a universal measure of the system's effective degrees of freedom, moreover, it provides important information about a given state, in particular, quantum correlations in a ground state between two spatially separated regions. For a system that decomposes into two subsystems, $\mathcal{A}$ and $\mathcal{B}$ with Hilbert space ${\cal H}={\cal H}_{\mathcal{A}}\otimes {\cal H}_{\mathcal{B}} $, the EE for a given quantum state $\vert \psi \rangle$ (e.g.  the ground state), is identified by the von Neumann entropy of the reduced density matrix \cite{Casini:2006es,Longo:2019pjj}                             \begin{equation}\label{EE}
		S_{EE}=-\text{Tr}( \rho_{\mathcal{A}} \log \rho_{\mathcal{A}}),
	\end{equation}
	where $\rho_{\mathcal{A}}=\text{Tr}_{\mathcal{B}}  (\vert \psi \rangle \langle \psi \vert)$ is 
	the reduced density matrix of the subsystem $\mathcal{A}$. 
	Despite the crucial role of EE in the study of quantum systems, by making use of the conventional methods in quantum field theories (QFTs), in general, there is no analytic solution for EE, though computing a reduced density matrix is an incredibly difficult task in many cases. Besides some standard methods for quantifying EE \footnote{For example,  the numerical methods which are mostly based on Gaussian states \cite{Casini:2009sr,Katsinis:2017qzh,Hertzberg:2010uv},  conformal field theory (CFT) methods \cite{Calabrese:2009qy,Ruggiero:2018hyl,Hung:2014npa} and holographic CFT \cite{holo,Ryu:2006ef}.}, for a quantum system described by a pure state, one may use the replica trick and make $n$ copies of
	system and define the  R\'enyi entropies as follows 
	 \cite{Headrick:2010zt}
	\begin{equation}\label{reny}
		S_{n}=\frac{1}{n-1}(n \log Z_1-\log Z_n),
	\end{equation}
	where in the path integral
	formalism, $Z_1$ and $ Z_n=Tr(\rho_{\mathcal{A}})^n$  are the partition functions on the original Riemann surface and on $n$-sheeted Riemann surfaces,  respectively. In this way, $S_n$ which carries all the information of the eigenvalues of the reduced density matrix can be computed  \cite{toni}. The $S_{EE}$ is then obtained, \begin{equation}
		S_{EE}=\lim_{n\rightarrow 1}\,S_n\,\,,
	\end{equation}
thus the computation of EE reduces to finding the partition functions.  On the other hand, the question of whether/how a certain symmetry of the system is related to the structure of entanglement has attracted attention in recent years
\cite{Goldstein:2017bua, german-3, Murciano:2020vgh,Capizzi-2020,Murciano-2021, Fraenkel:2019ykl,Murciano:2020lqq,Azses:2020wfx, Parez:2022sgc}. 
 In fact, besides the theoretical motivation, there is also experimental motivation to investigate such a relation
  between entanglement and underlying symmetry of a given system, e.g., in Ref. \cite{Lukin19}, the authors explored the relation between the dynamics of many-body disordered systems and the entanglement that is formed from various symmetry sectors. 
	The idea of making a connection between a certain amount of entanglement and specific individual symmetry sectors of a quantum theory is known as the symmetry resolution of entanglement. Particularly, SREE is a theoretical framework to understand the role of the contribution of different symmetry sectors in computing the EE  \cite{Goldstein:2017bua, german-3}.  
It should be noted that besides the method of SREE, many advances have been made to investigate other features of the effect of symmetry in computing entanglement, for example, symmetry-resolved relative entropy \cite{Chen:2021pls, Capizzi:2021zga}, symmetry decomposition of entanglement negativity \cite{Cornfeld:2018wbg, Chen:2021nma, Parez:2022xur, Gaur:2022sjf}, SREE in holographic settings \cite{Zhao:2020qmn, Weisenberger:2021eby, Zhao:2022wnp} in thermal states \cite{Ghasemi:2022jxg, Piroli:2022ewy}, excited states \cite{Capizzi:2022jpx, Capizzi:2022nel}, free lattice models \cite{Bonsignori:2019naz, Jones:2022tgp}, integrable field theories \cite{Horvath:2020vzs,Horvath:2021fks,Horvath:2021rjd} and SREE after a quench \cite{Parez:2020vsp,Scopa}.  \\ 
In this paper, we employ the correlator method to determine the eigenvalues of the correlation matrix and subsequently compute the symmetry-resolved R'enyi entropies for free complex scalar quantum fields and a specific class of non-local QFTs introduced in \cite{Shiba:2013jja}. To study the SREE, we demonstrate that by identifying the four largest eigenvalues for the massless limit and the two largest eigenvalues for the massive limit, it is possible to derive effective expressions for the eigenvalues of the correlation matrix in both regimes, thereby enhancing the efficiency of numerical analysis. Additionally, we investigate the equipartition of entanglement across different charge sectors, confirming the effective equipartition of entanglement for massless cases and for non-local QFTs.
\\ The organization of this paper is as follows: Section 2 briefly recalls the SREE and the correlator method used for the numerical analysis, Section 3 studies the SREE and the related equations for complex scalar fields, Section 4 is devoted to the numerical analysis of the expressions obtained in Section 3, and Section 5 focuses on the study of EE and SREE for non-local QFTs. Concluding remarks are presented in Section 6, and some details of computing symmetry-resolved partition functions for a single site are provided in Appendix A.

	
	\section{SREE and correlator method
	}
	\label{sec:SR}
	In the context of global symmetry, the total EE can be decomposed into symmetry sectors. Consider a system with a global internal symmetry that divides into two complementary spatial subsystems, $\mathcal{A}$ and $\mathcal{B}$. In a particular state described by the density matrix $\rho$, the symmetry is generated by a charge operator ${Q}={Q}_{\mathcal{A}}\oplus {Q}_{\mathcal{B}}$. If the state is an eigenstate of the symmetry generator ${Q}$, then $[\,{\rho}\,,\,{Q}\,]=0$. Tracing over $\mathcal{B}$ results in $[\,{\rho}_{\mathcal{A}},\,{Q}_{\mathcal{A}}] = 0$. Consequently, ${\rho}_\mathcal{A}$ is block-diagonal, with each block corresponding to a different charge sector and labeled according to the eigenvalue $q$ of the charge operator ${Q}_{\mathcal{A}}$, as follows:
	\be
	\label{TEE}
	\rho_\mathcal{A}=\bigoplus_{q} p(q)\rho_{\mathcal{A}}(q),
	\ee
	where  $ p(q) = \text{Tr}(\Pi_q \rho_\mathcal{A})$ is the probability that the measurement of $Q_\mathcal{A}$ in region $\mathcal{A}$ is $q$, and  $\Pi_q$ stands for the projector on the sector of $Q_\mathcal{A}$ with eigenvalue $q$.   
	The EE at charge $q$ which is defined according to $\rho_\mathcal{A}$ is then given by 
	\begin{equation}
		\label{eq:SvNEE}
		S(q) = -\mathrm{Tr}\big (\rho_\mathcal{A}(q) \log \rho_\mathcal{A}(q)\big),
	\end{equation}
	that is known as the SREE. 
	According to \eqref{TEE}, the total EE associated with the density matrix $\rho_A$ can be decomposed as,
	\begin{equation}
		\label{Eq:SvN}
		S_{EE}=\displaystyle \sum_q p(q) S(q)- \displaystyle \sum_q p(q) \log p(q),
	\end{equation}
	where the first part of the right-hand side of the above equation is known as the configurational entropy that denotes the average entropy per sector; the second part is the number entropy stands for the charge fluctuations within the subsystem $\mathcal{A}$ \cite{Lukin19}. Similarly, one may use the replica trick and define the symmetry-resolved R\'enyi entropies 
	\begin{equation}
	\label{Eq:SReen}
	S_n(q)=\dfrac{1}{n-1}\Big(n\log  \mathcal{Z}_1(q)-\log\mathcal{Z}_n(q)\Big), 
	\end{equation}
	where $\mathcal{Z}_n(q)= \mathrm{Tr} (\Pi_{q}\,\rho^n_\mathcal{A})$ is the symmetry-resolved partition function. To determine these partition functions, it is necessary to simultaneously diagonalize the reduced density matrix and $Q_\mathcal{A}$, a task that is generally challenging. The key concept in the symmetry resolution of entanglement is to employ the path integral approach to facilitate the computation of $\mathcal{Z}_n(q)$ \cite{Goldstein:2017bua}. 
In this formalism, an Aharonov-Bohm flux $\alpha$ is introduced on each sheet of the Riemann geometry, resulting in the field acquiring a total phase $\alpha$ after considering the entire Riemann surface $\mathcal{R}_{n}$. This introduces a new quantity in the calculation of the path integral, known as the charged moments of the reduced density matrix, defined by:  
	\begin{equation}
		\label{eq:firstdef}
		Z_n(\alpha)=\mathrm{Tr}(\rho_\mathcal{A}^n\,\,e^{iQ_\mathcal{A} \alpha}).
	\end{equation}
It is important to highlight that the path integral method offers a structure for determining $Z_n(\alpha)$ by tracking the evolution of the partition function of an $n$-copy quantum field based on the given boundary condition \cite{Belin:2013uta}.	The Fourier transforms of the charged moments give us the symmetry-resolved partition function
		\begin{equation}\label{Eq:defZnq}
		\mathcal{Z}_n(q)=\displaystyle \int_{-\pi}^{\pi}\dfrac{d\alpha}{2\pi}e^{-iq\alpha}\,Z_n(\alpha).
	\end{equation} 
Finally, the SREE can be obtained as follows
\bea 	\label{Eq:SReen1}
S(q)=\lim_{n\rightarrow 1} S_n(q)=\log\mathcal{Z}_1(q) -\frac{\left.\big(\partial_{n}\mathcal{Z}_n(q)\big)\right|_{n=1}}{\mathcal{Z}_1(q)},
\eea
it should be noted that the probability $p(q)$ is simply related to the moments $\mathcal{Z}_n(q)$ as follows
\begin{equation}
p(q)=\mathcal{Z}_1(q).
\end{equation}		
This study focuses on a free and quadratic theory, utilizing the correlator method for numerical computation. The method was initially presented in \cite{JP}, and for further information, refer to \cite{Doroudiani:2019llj, Mollabashi:2020ifv, Ghasemi:2021jiy,  MohammadiMozaffar:2017nri,Khorasani:2023usq}. The approach is employed to determine the charged moments and symmetry-resolved partition functions.
\subsection{Correlator method}

For Gaussian states, the numerical computation of EE primarily relies on the two-point function of degrees of freedom within a specified region of the subsystem. A similar approach is employed for SREE. However, one must first evaluate the charged moments using the correlation or two-point functions. This is followed by computing the symmetry-resolved moments, and finally, constructing the symmetry-resolved R\'enyi and entanglement entropies. For a given free theory, the correlator method which works for Gaussian states provides a powerful numerical framework to find the eigenvalues of the correlation matrix e.g., for free scalar field theories one should compute the following two-point functions
\be
\frac{1}{2}\mathbf{X}_{ij}=\langle \Phi_i \Phi_j \rangle\,,\;\;\;\;\;\;\;\;\;\;\frac{1}{2}\mathbf{P}_{ij}=\langle \Pi_i \Pi_j \rangle.
\ee
where $\Pi(x)$ is the momentum conjugated of $\Phi(x)$. In this method, for a subregion $\mathcal{A}$, the correlation matrix $\Lambda_\mathcal{A}= \sqrt{\mathbf{X}_\mathcal{A} \cdot \mathbf{P}_\mathcal{A}}$ should be constructed, where its eigenvalues are utilized to determine the EE. Specifically, it has been shown that in the case of a bipartition, where the subsystem $\mathcal{A}$ consists of $\ell$ adjacent lattice sites, the EE for Gaussian states is given by:	\begin{equation}\label{Eq:SEE}		S_{EE}=\sum_{i=1}^{\ell}\bigg[\left(\frac{\nu_i+1}{2}\right)\log\left(\frac{\nu_i+1}{2}\right)- \left(\frac{\nu_i-1}{2}\right) \log\left(\hspace{.5mm}\frac{ \nu_i-1}{2}\hspace{.5mm}\right)\bigg], 
	\end{equation} where each $\nu_i$ represents the eigenvalues of the matrix $\Lambda_\mathcal{A}$. This approach is broadly applicable to quadratic free quantum field theories. As previously mentioned, we further elaborate on this technique to determine SREE for complex scalar fields.


	\subsubsection{ Free bosonic quantum many-body systems}\label{sec:harmonic}
Gaussian states are fully characterized by the correlation functions of the canonical variables, which can be employed to calculate quantum entanglement. Conversely, harmonic lattice systems can be viewed as discrete analogs of Klein-Gordon fields. For lattice discretization of a given Hamiltonian with periodic boundary conditions, the continuous spatial coordinates $x$ are replaced by a lattice of discrete points with $\mathcal{N}$ sites and lattice spacing $\epsilon$. For simplicity, we have set $\epsilon = 1$. The  Hamiltonian can be written in the following form 
\begin{equation}
{\cal H}= \sum_{r=1}^{\mathcal{N}} \dfrac{1}{2} \mathcal{P}_{r}^2 + \sum_{r,s=1} ^{\mathcal{N}}\dfrac{1}{2} \mathcal{Q}_{r} K_{rs} \mathcal{Q}_{s} , \label{eq:2-1}
\end{equation}
where $K$ is a symmetric and positive-definite matrix, the displacement and momentum conjugation of the  $i$-th oscillator are determined by $ \mathcal{Q}_i$ and $ \mathcal{P}_i$. The Gaussian wave function of the ground state is given by 
	\begin{equation}
		\Psi (\{\mathcal{Q}\})=\left( \det \dfrac{W}{\pi}\right)^{1/4} \exp \Big[-\frac{1}{2}\sum_{r,s}\mathcal{Q}_r W_{rs}  \mathcal{Q}_s\Big], \label{eq:1-2}
	\end{equation}
	where  $ W \equiv K^{1/2} $.
After making use of the discrete Fourier transformation, the diagonalized $ K $ matrix yields as follows 
	
	\begin{equation}
		K_{rs}= \frac{1}{ \mathcal{N}} \sum_{k} \Big[ m^2+2  (1-\cos \dfrac{ 2\pi k }{\mathcal{N}} ) \Big] e^{2\pi i k(r-s)/\mathcal{N}}. \label{eq:2-3}
	\end{equation}
	In order to have the limit of a large chain (with fixed   $\epsilon$), we need to take $ \mathcal{N} $ to infinity and $p=2\pi k /\mathcal{N}$. The correlation functions are given by
	
	\begin{equation} \label{Eq:towpointPKG}
		\mathbf{P}_{rs}= \int_{-\pi}^{\pi} \dfrac{d p}{(2\pi )} e^{i p (r-s)}\sqrt{ m^2 + 2 (1-\cos p ) }  
	\end{equation}
	
	\begin{equation} \label{Eq:towpointQKG}
		\mathbf{X}_{rs}= \int_{-\pi}^{\pi} \dfrac{d p}{(2\pi )} e^{i p (r-s)} \frac{1}{ \sqrt{m^2 + 2  (1-\cos p )} } 
	\end{equation}
Substituting \eqref{Eq:towpointPKG} and \eqref{Eq:towpointQKG} into the $\Lambda_\mathcal{A}$ allows for the determination of the eigenvalues of the correlation matrix. As a result, equation \eqref{Eq:SEE} provides us with the EE. It is worth noting that numerical methods are frequently employed in the literature to compute the integrals mentioned above. We have come across a useful relation that may be of interest
	\begin{equation}\label{Eq:anaPQKG}
		\int_{-\pi }^{\pi } \frac{\cos (\mathfrak{n} p) \Big[m^2+\big[2-2 \cos (p)\big]\Big]^{\mathfrak{j}}}{2 \pi } \, d p=\left(m^2+4\right)^\mathfrak{j} \, _3\tilde{F}_2\left(\frac{1}{2},1,-\mathfrak{j};1-\mathfrak{n},1+\mathfrak{n};\frac{4}{m^2+4}\right),
	\end{equation}
	where $\mathfrak{n} $  is  an integer number and ${_3\tilde{F}_2 }$  is the regularized hypergeometric function. 
	

	\section{SREE for complex bosonic theory }
	\subsection{Complex harmonic chain}
Field theory offers a more effective approach to examining the symmetry of harmonic chains and their lattice discretization. In order to explore the characteristics of complex harmonic chains, the complex scalar field is commonly employed \cite{Murciano:2019wdl, Castro-Alvaredo:2018dja, Castro-Alvaredo:2018bij, Capizzi:2022nel}.
	The Euclidean action and  the corresponding Hamiltonian of a free complex massive scalar field $\Phi(x)$ are given by 
	\bea
	{\cal I}&=&\int d^2 x \left[ \partial_{\mu} \Phi^\dag(x)\partial_{\mu} \Phi(x) +m^2 \Phi^\dag(x)\Phi(x)\right],\nonumber\\
	{\cal H}&=&\int d x \left[\Pi^\dag(x)\Pi(x)+\partial_x \Phi^\dag(x)\partial_x \Phi(x) +m^2\Phi^\dag(x) \Phi(x)\right],
	\label{hc}
	\eea
	and it is shown that the fields satisfy the following commutation relations
	\be
	[\Phi(x),\Pi(y)] = i\delta(x-y) \,.
	\ee
	In terms of creation and annihilation operators, one can write
	\begin{equation}\label{Eq:Phi} 		\Phi=\int\frac{d p}{2 \pi}\frac{1}{\sqrt {2 \omega}}\,(a(p) e^{ipx}+b^\dagger(p) e^{-ipx}),\,\,\,\,\,
		\Pi=\int\frac{d p}{2 \pi}i\sqrt\frac{ \omega}{2}\,(a^\dagger(p) e^{-ipx}- b(p) e^{+ipx}),	\end{equation}
		note that  $\omega^2(p)=m^2+p^2$. Understanding a specific symmetry involves recognizing the action's invariance. For instance, when rotating the field $\Phi$ by an arbitrary phase $\Phi(x)\to e^{i\theta}\Phi(x)$, the action \eqref{hc} stays the same, indicating the presence of $U(1)$ symmetry. To make the $U(1)$ symmetry manifest, one may write the $\Phi(x)$ in terms of two real scalar fields, $\Phi_1(x)$ and $\Phi_2(x)$ as  
	\begin{equation}
		\Phi(x)=\frac1{\sqrt2} (\Phi_1(x)+ i\Phi_2(x))\,, \hspace{4mm}\Pi(x)=\frac1{\sqrt2} (\Pi_1(x)+ i\Pi_2(x)),
	\end{equation}
thus the Hamiltonian of the complex scalar fields  decomposes into two real Hamiltonian as follows
	\be {\cal H}={\cal H}(\Phi_1)+{\cal H}(\Phi_2)
	=\frac12\sum_{i=1}^{2}\int d x \left[\Pi_i^2(x)+(\partial_x \Phi_i(x))^2 +m^2\Phi_i^2(x)\right] 
			\label{Eq:H2real},
		\ee
it is then easy to verify that the $U(1)$ turns to the $O(2)$ rotation. The Hamiltonian and the conserved charge can be rewritten in terms of the creation and annihilation operators 
	\begin{equation}
		{\cal H}=\int \frac{d p}{2\pi} \omega(p) \Big(a^\dag(p) a(p) +b^\dag(p) b(p) \Big),\,\,\,\,\,\,\,
		Q=\int \frac{d p}{2\pi}   \Big(a^\dag(p) a(p) -b^\dag(p) b(p)\Big ),
	\end{equation}
	where \begin{equation}
		a(p)=\frac1{\sqrt2} (a_1(p) +i a_2(p)), \qquad b(p)=\frac1{\sqrt2} (a_1^\dag(p) +i a_2^\dag(p)).
		\label{ab:op}
	\end{equation}
	Now in real space, by restriction of the domain of integration into the subregion ${\cal A}$, one can obtain the conserved charge in ${\cal A}$ which is given by 
	\begin{equation}
		Q_{\cal A}=\int_{\cal A} d x  \Big(a^\dag(x) a(x) -b^\dag(x) b(x) \Big).
		\label{QAft}
	\end{equation}
To proceed with the numerical method, it is necessary to discretize the theory, a process that involves creating a lattice representation of the complex scalar field.
	\begin{figure}[h]
	
	\centering
	\includegraphics[scale=.29]{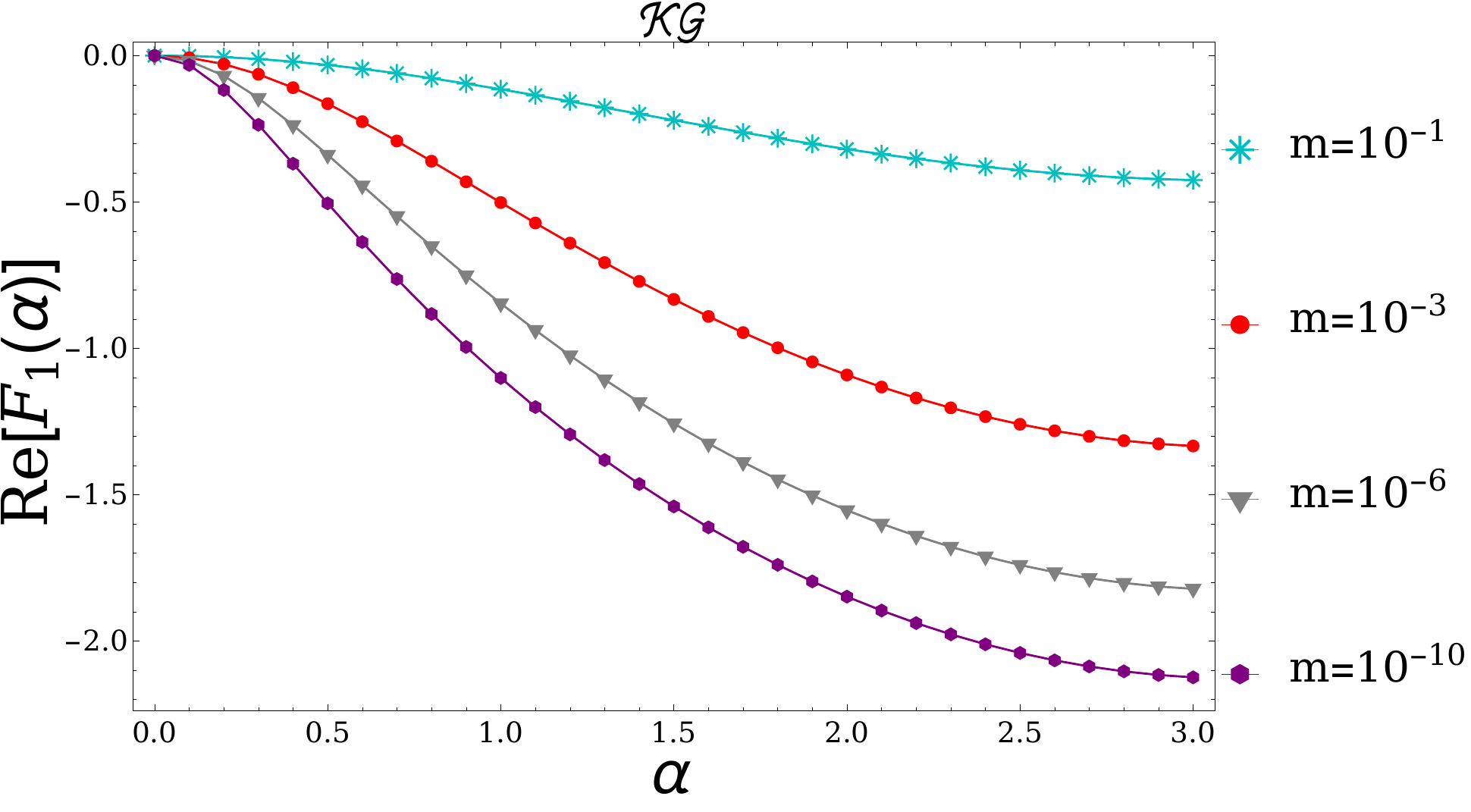}\includegraphics[scale=.29]{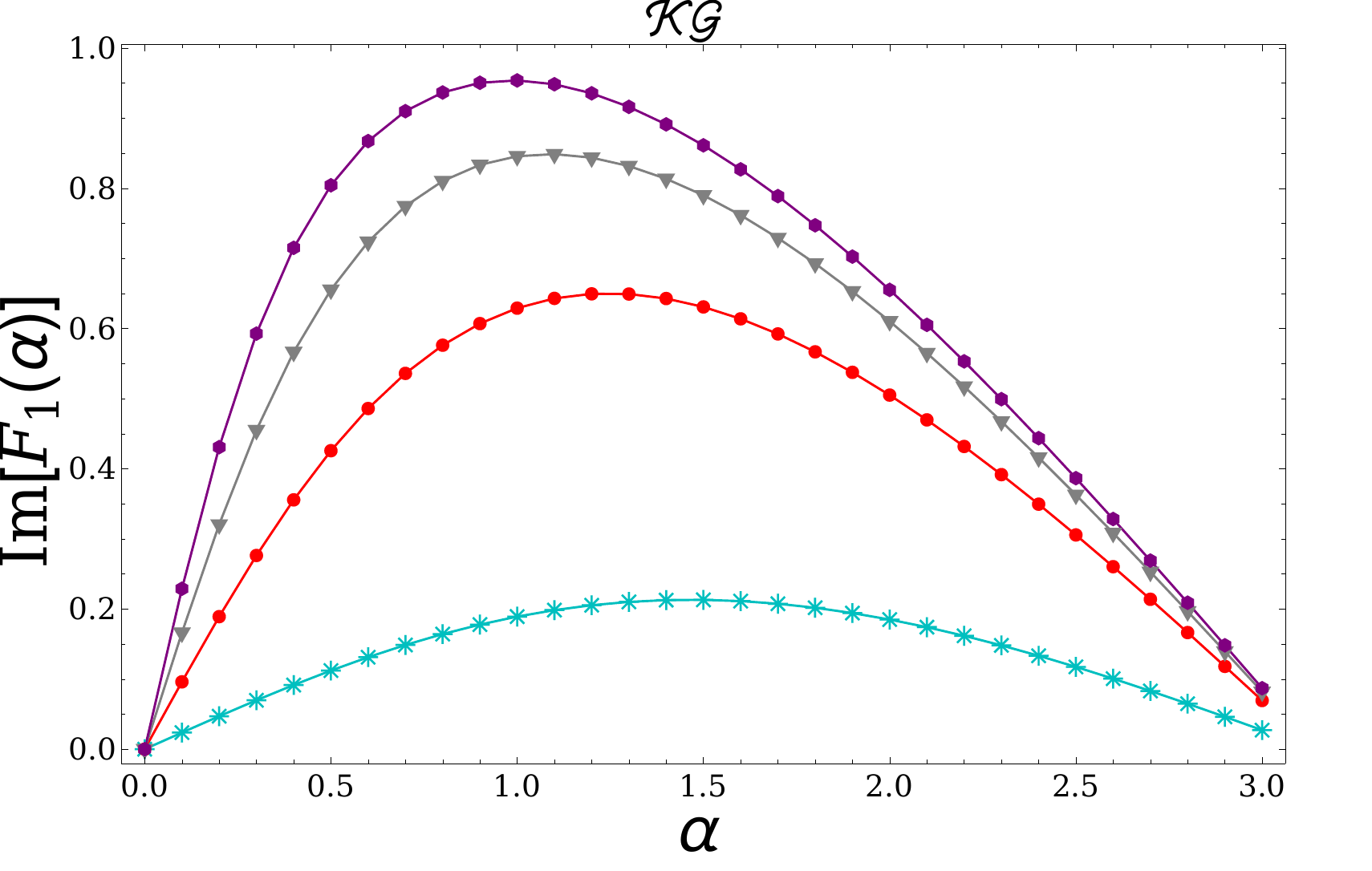} 
	\includegraphics[scale=.29]{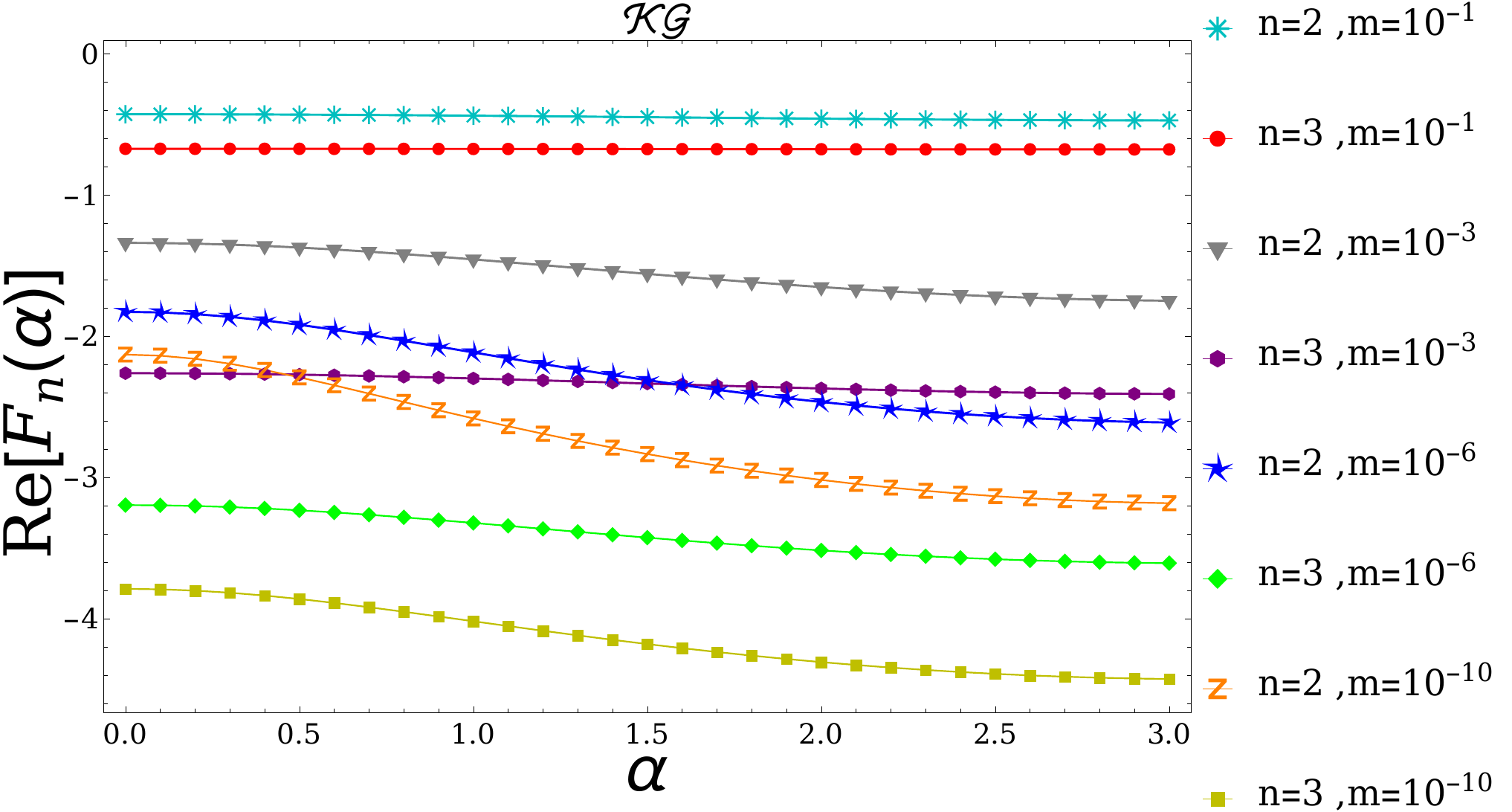}\includegraphics[scale=.29]{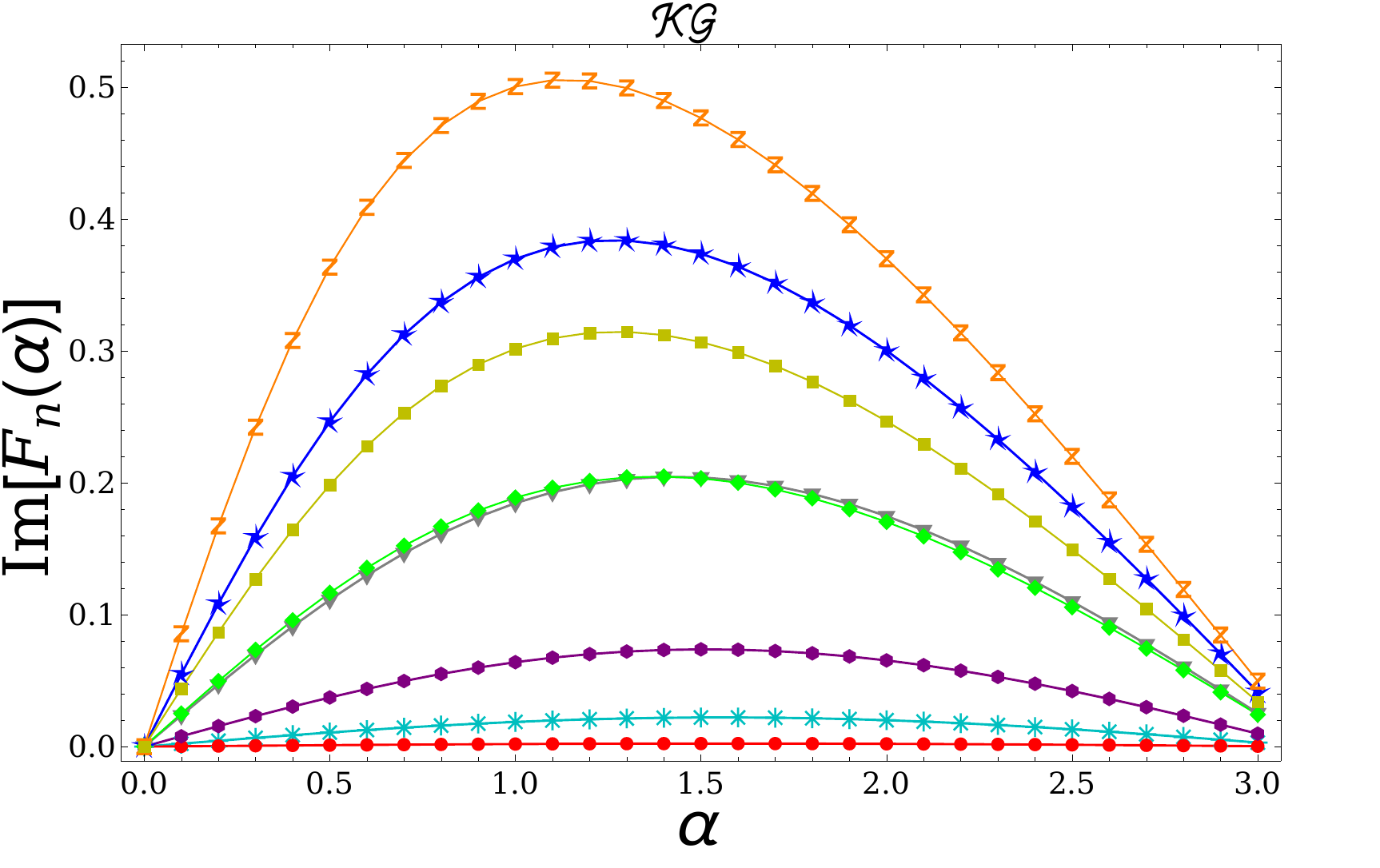} 	 
	\caption{ The $F_n(\alpha)$ function's numerical results for the real and imaginary parts are displayed with respect to $\alpha$, with $\ell=70$ and $m=10^{-1}, 10^{-3}, 10^{-6}, 10^{-10}$, and for $n=1$ (top plot), $n=2$, and $n=3$ (bottom plot). } \label{Fig:FnalphanotcnstKG}
\end{figure}
\begin{figure}[h]
	\centering
	\includegraphics[scale=.28]{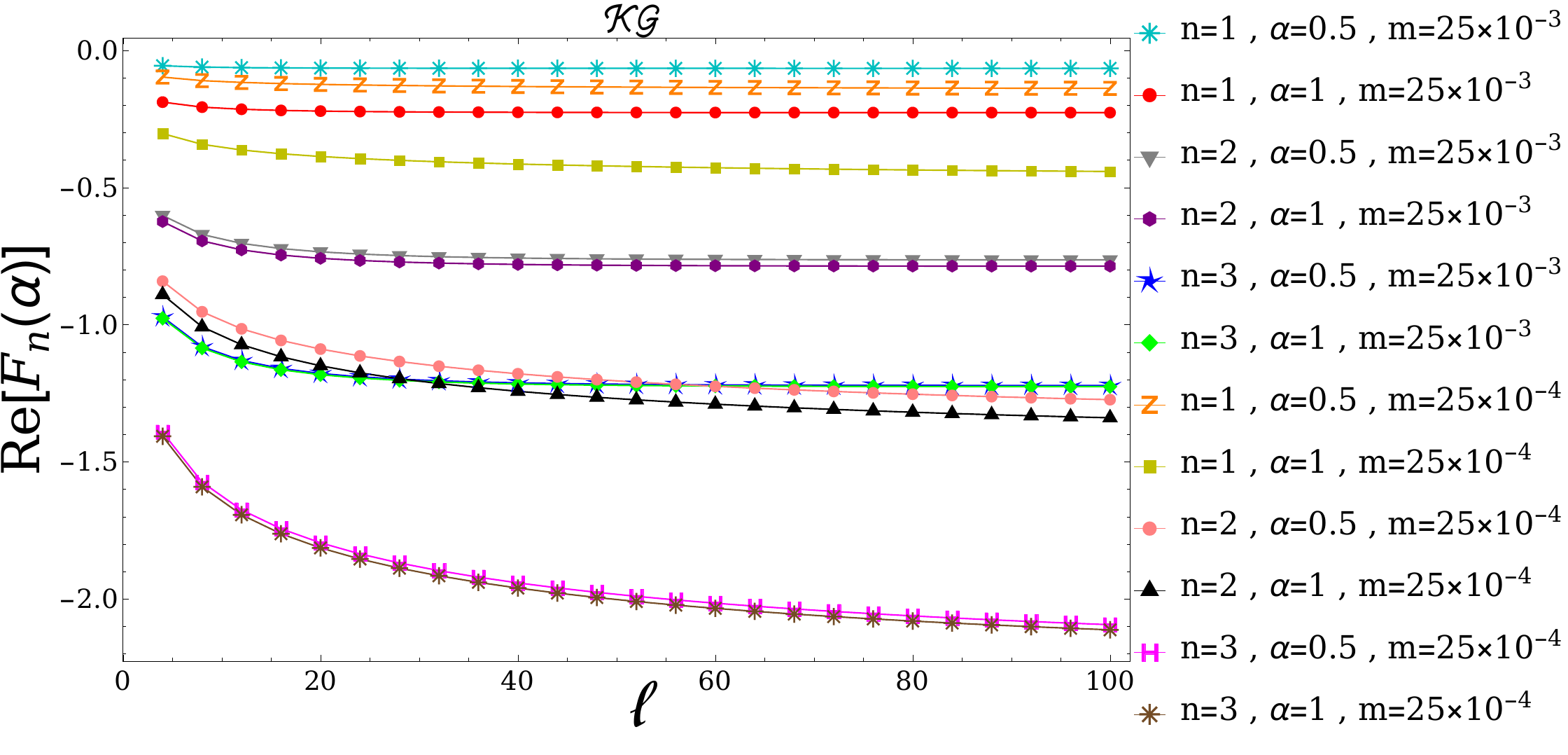}\includegraphics[scale=.28]{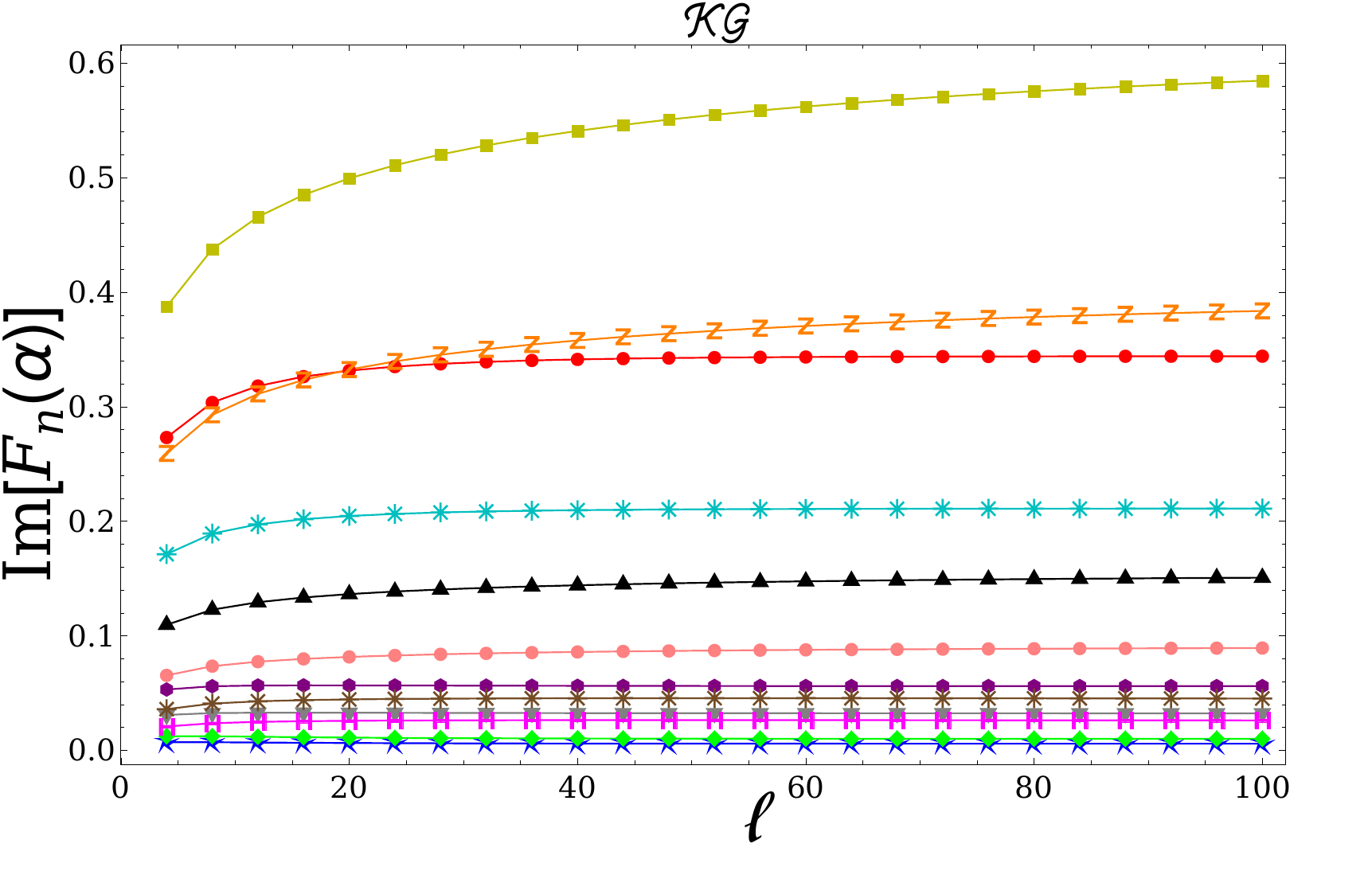} 	 
	\caption{ Numerical results of the real and imaginary parts of the $F_n(\alpha)$ function are shown in terms of $\ell$, for $m=25 \times 10^{-3}$ and  $m=25 \times 10^{-4}$ and  $\alpha =0.5$ and $1$  for $n=1, 2$ and $3$. } \label{Fig:FNalphacnst}
\end{figure}
\begin{figure}[h]
	\centering
	\includegraphics[scale=.47]{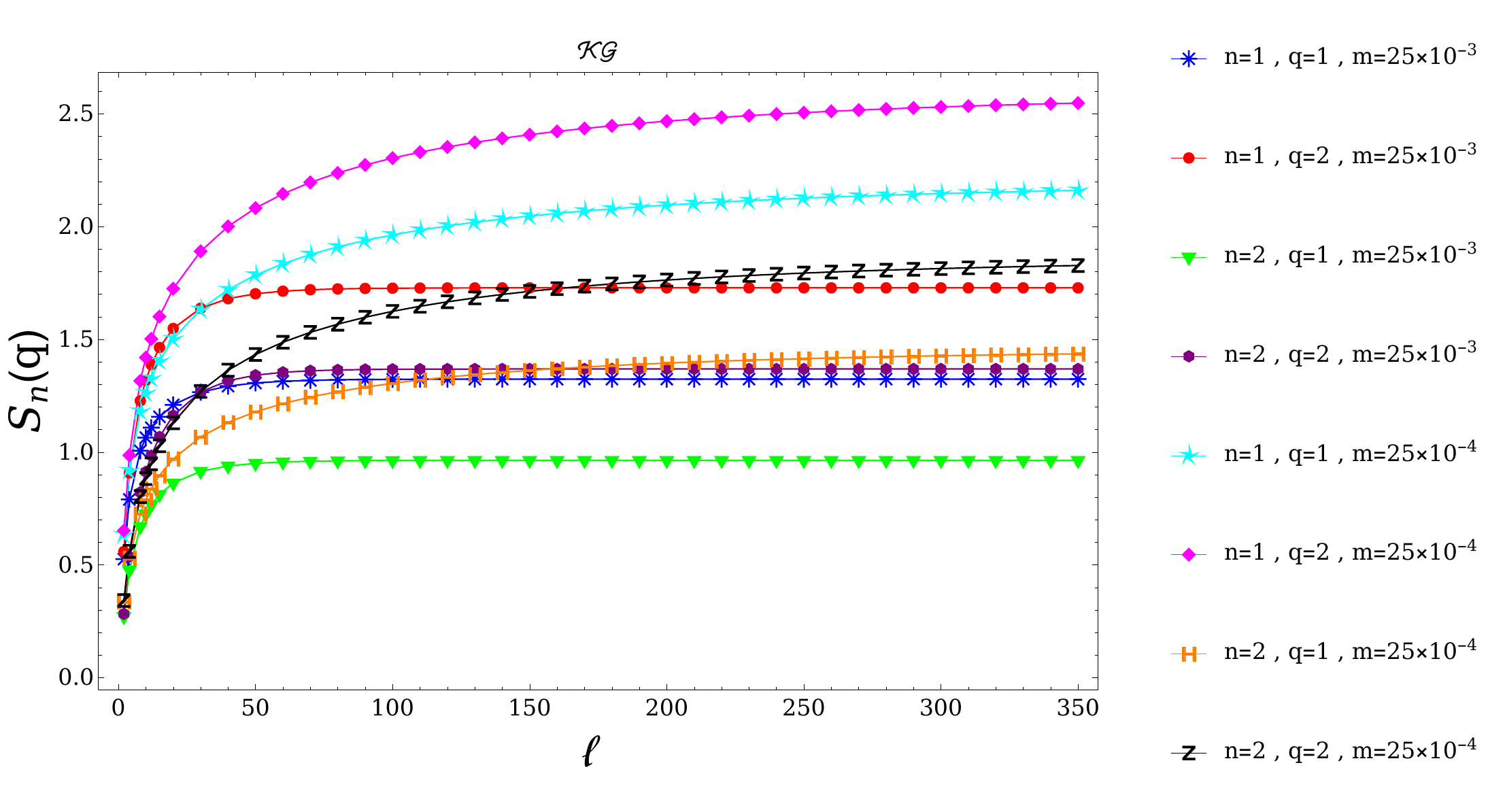} 	 
	\caption{The numerical values of $S_n(q)$ for the complex harmonic chain are shown in terms of $\ell$ where we set $m=25 \times 10^{-3}$ and $~m=25 \times 10^{-4} $,  $n=1,\,2$ and $q=1,\,2$. For definite given values of $n$ and $q$, 
	it can be seen that $S_n(q)$ converges at smaller values of $\ell$ with larger mass.} \label{Fig:SNQKG}
\end{figure}
\begin{figure}[h]
	\centering
	\includegraphics[scale=.28]{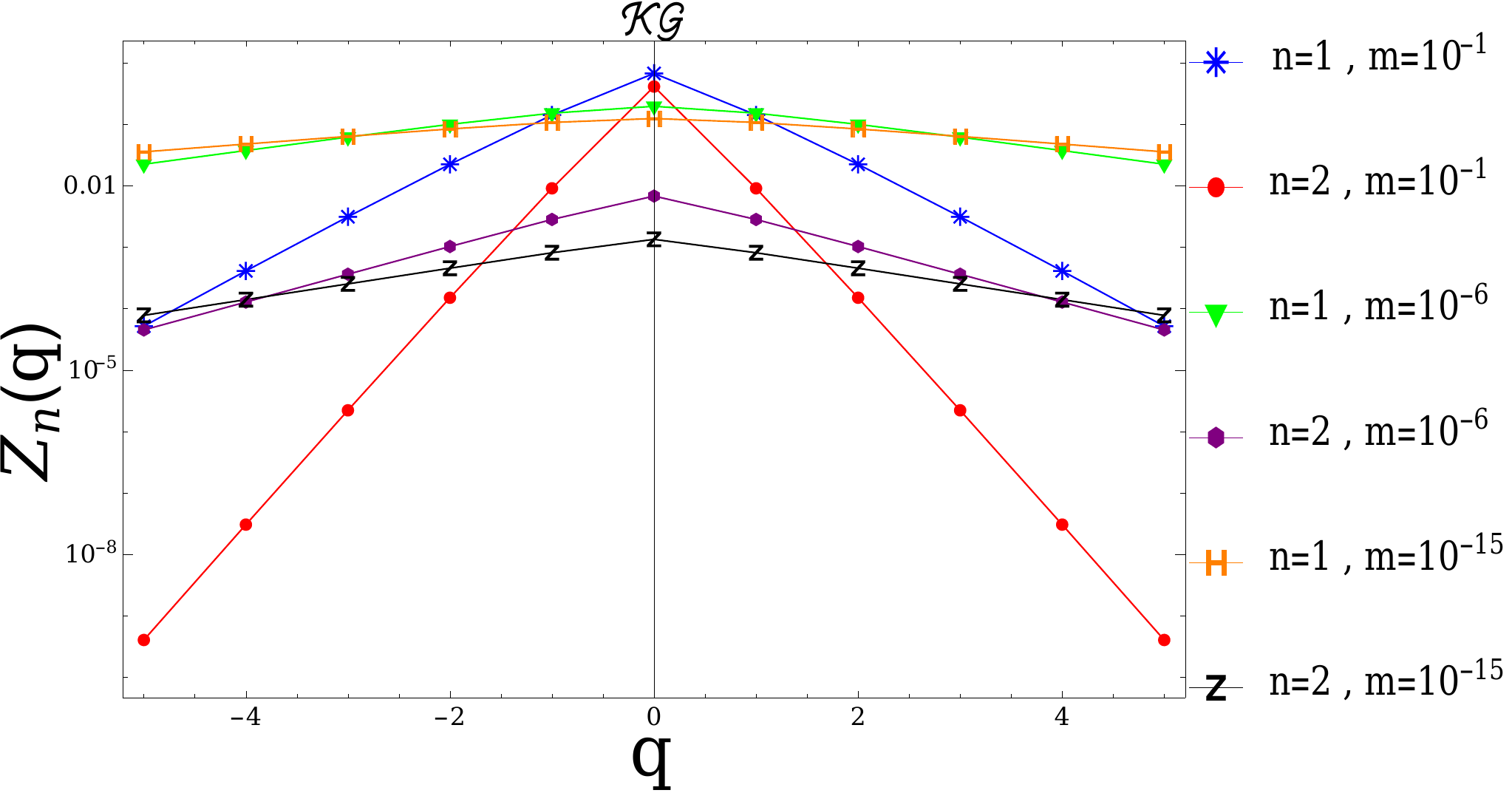}\includegraphics[scale=.28]{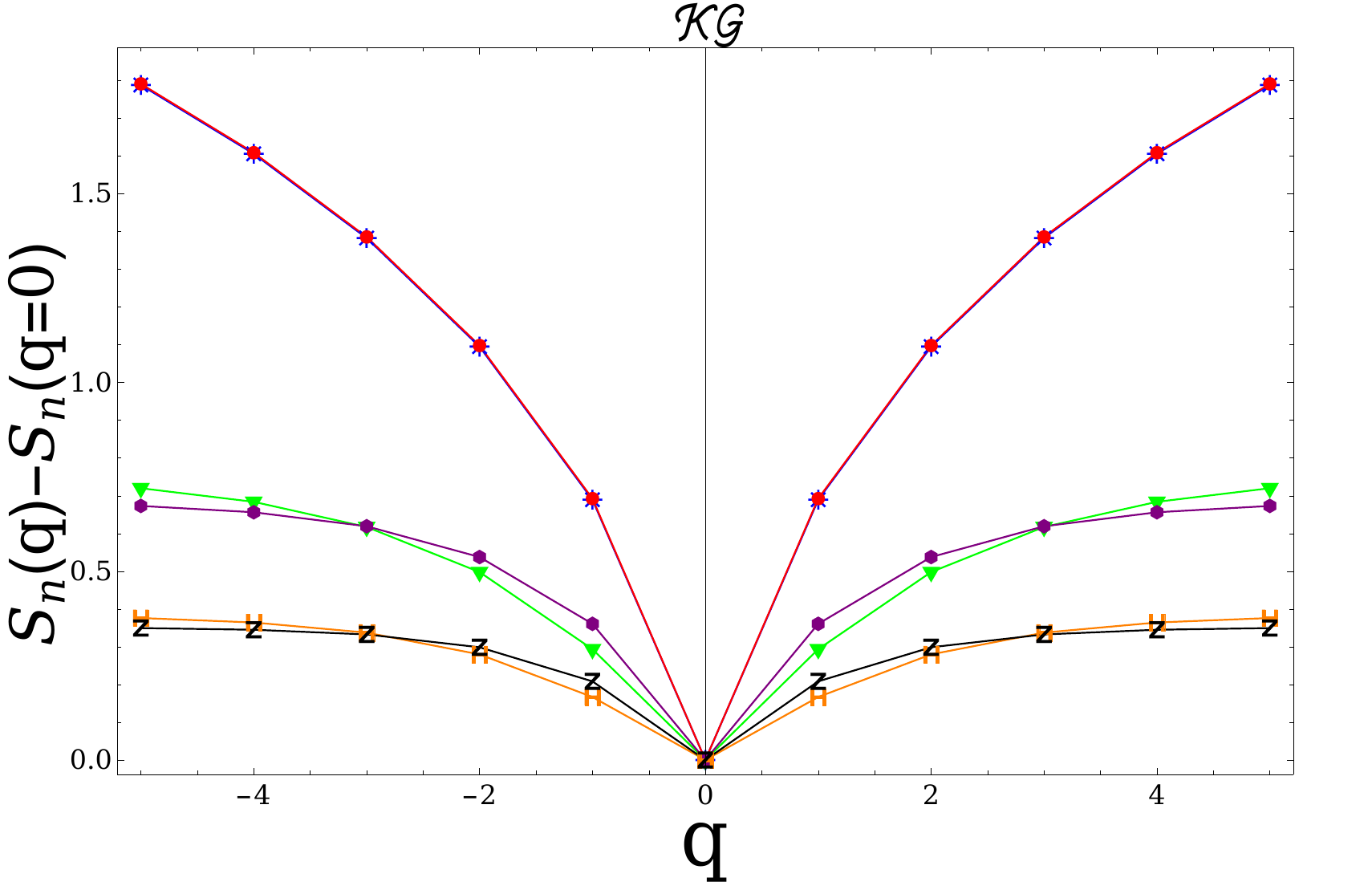} 	 
	\caption{ The numerical results for $\mathcal{Z}_n(q)$ and $ S_n(q)$ in terms of $q$, for $m= 10^{-1}$, $m= 10^{-6}$ and $m= 10^{-15}$, we set $\ell=250$  and $n=1$ and $2$. For larger values of mass, the left panel shows the $q$-dependence of symmetry-resolved moments, while this dependence is relaxed in the light masses. This behavior also holds for $S_n(q)$, as can be seen from the right panel, the dependence $S_n(q)$ on $q$ becomes more relaxed for light masses and larger value of $q$. } \label{Fig:ZnSqchange}
\end{figure}
	\subsection{The lattice model}
	
	In order to maintain a continuous symmetry, we can examine a complex bosonic theory that can be represented on the lattice as a chain of complex oscillators. Alternatively, we can view it as the combination of two real harmonic chains involving the variables $(\mathcal{P}^{(1)},\mathcal{Q}^{(1)})$ and $(\mathcal{P}^{(2)},\mathcal{Q}^{(2)})$. The corresponding Hamiltonian is given by
	\begin{equation}
		\label{eq:complex}
		{\cal H}_{CB}(\mathcal{P}^{(1)}+i\mathcal{P}^{(2)},\mathcal{Q}^{(1)}+i\mathcal{Q}^{(2)})={\cal H}_{B}(\mathcal{P}^{(1)},\mathcal{Q}^{(1)})+{\cal H}_{B}(\mathcal{P}^{(2)},\mathcal{Q}^{(2)}).
	\end{equation}
	The Hamiltonian (\ref{eq:complex}) can be also written
	in terms of mode operators $a_k$ and $b_k$ satisfying $[a_k,a^{\dagger}_j]=\delta_{j,k}$, $[b_k,b^{\dagger}_j]=\delta_{j,k}$, as follows 
	\begin{equation}
		{\cal H}_{CB}=\sum_{k=0}^{L-1} \omega_k(a^{\dagger}_ka_k+b_{k}^{\dagger}b_k), \qquad \omega_k=\sqrt{ m^2+2  (1-\cos \dfrac{ 2\pi k }{\mathcal{N}} )},
	\end{equation}
	and similarly, the charge operator reads
	
	\begin{equation}
		Q=\sum_{k=0}^{L-1}\Big(a^{\dagger}_ka_k- b_{k}^{\dagger}b_k\Big).
	\end{equation}
The conserved quantity can be effectively expressed in real space, and its value within a specific subsystem $\mathcal{A}$ is equal to the sum confined to $\mathcal{A}$, namely
		\begin{equation}
		\label{eq:chargeanti}
		Q_\mathcal{A}=\displaystyle \sum_{j\in \mathcal{A}} \Big( a^{\dagger}_ja_j-b^{\dagger}_jb_j\Big).
	\end{equation}
In order to calculate the charged moments, it is necessary to evaluate ${\rm Tr} [ \rho_\mathcal{A}^n e^{i {Q}_\mathcal{A}\alpha} ]$. By utilizing equation \eqref{eq:chargeanti} for ${Q}_\mathcal{A}$, the trace can be factorized as
	\begin{equation}
		Z_n(\alpha)={\rm Tr}[\rho_\mathcal{A}^n e^{i {Q}_\mathcal{A}\alpha}]= {\rm Tr}[(\rho_\mathcal{A}^{a})^n e^{iN_\mathcal{A}^{a}\alpha}]\times {\rm Tr}[(\rho_\mathcal{A}^{b})^n e^{-iN_\mathcal{A}^{b}\alpha}],
	\end{equation}
	where $N_\mathcal{A}^{a}=\sum_{j\in A} a^{\dagger}_{j}a_{j}$ and 
	$N_\mathcal{A}^{b}=\sum_{j\in \mathcal{A}} b^{\dagger}_{j}b_{j}$. Using the relations between the number operator $N_\mathcal{A}$ and the eigenvalues of the correlation matrix $\nu_k$, we find 
	
	\begin{equation}\label{Eq:Znalpha}
		Z_n(\alpha)=
		\prod_{k=1}^\ell
		\frac{2^{n}}{
			\left(
			\nu_k+1
			\right)^n
			-
			e^{i\alpha}
			\left(
			\nu_k-1
			\right)^n
		}
		\frac{2^{n}}{
			\left(
			\nu_k+1
			\right)^n
			-
			e^{-i\alpha}
			\left(
			\nu_k-1
			\right)^n
		},
	\end{equation}
	and by the following definition for a single harmonic chain \cite{Murciano:2019wdl , Murciano:2020vgh}
	\begin{equation}\label{Eq:Fn}
		F_n(\alpha)\equiv\log {\rm Tr}[(\rho_A^{a})^n e^{iN_A^{a}\alpha}].
	\end{equation}
	one can write 
	\begin{equation}\label{Eq:LogZn}
		 Z_n(\alpha)=\exp\Big[F_n(\alpha)+F_n(-\alpha)\Big].
	\end{equation}
In the upcoming section, through numerical analysis, we will utilize the aforementioned formula to calculate $\mathcal{Z}_n(q)$, and the SREE can be acquired using equation \eqref{Eq:SReen}.

	\section{Numerical analysis }
In this section, we employ numerical methods to compute the EE for complex scalar fields and analyze its partition into different charge sectors for lattice discretization. In the subsequent section, we apply the same methodology to non-local fields. For the complex scalar fields, we build upon the results from the previous section. By elaborating on the $F_n(\alpha)$ functions in Figure \ref{Fig:FnalphanotcnstKG}, we plot $F_n(\alpha)$ as a function of $\alpha$. Additionally, in Figure \ref{Fig:FNalphacnst}, we present the numerical results of $F_n(\alpha)$ as a function of $\ell$ for two different masses and specific values of $\alpha$ and $n$. The figures are consistent with the similar results reported
in  Ref. \cite{Murciano:2019wdl}, where an analysis based on corner transfer matrix was employed to investigate $F_n(\alpha)$ and $S_n(q)$. In the graph presented in Figure \ref{Fig:SNQKG}, $S_n(q)$ is plotted as a function of $\ell$  for specific values of  of $q$, $n$, and mass. It is observed that the behavior of $S_n(q)$ is influenced by mass, resulting in saturation at lower values of $\ell$ for heavier masses. For further exploration, the behavior of $\mathcal{Z}_n(q)$ and $ S_n(q)$ for various mass values is illustrated in Figure \ref{Fig:ZnSqchange}. The graphs reveal that for higher masses, the correlation with $q$ becomes evident, as both exhibit noticeable variations with $q$. Nevertheless, the 
 changes (slope) in graphs decrease for lighter masses, indicating that the $q$-dependence of the symmetry-resolved moments becomes smaller for light masses. This pattern is also observed for $S_n(q)$, suggesting that in this regime, the equipartition of the EE is effectively established.

	\subsection{Massless and massive limit}\label{sub-mass}
\begin{figure}[h]
		\centering
		\includegraphics[scale=.32]{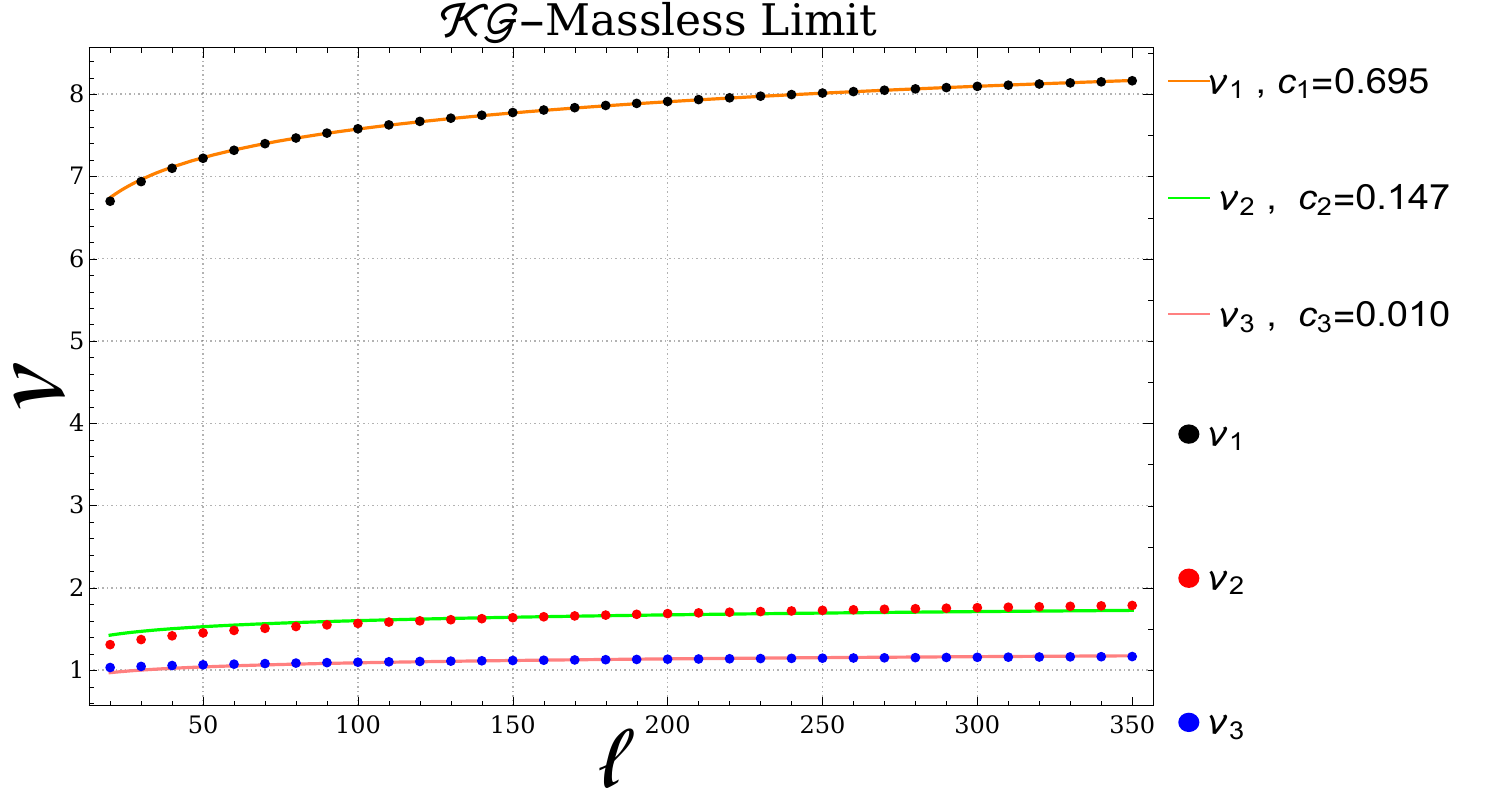} \includegraphics[scale=.32]{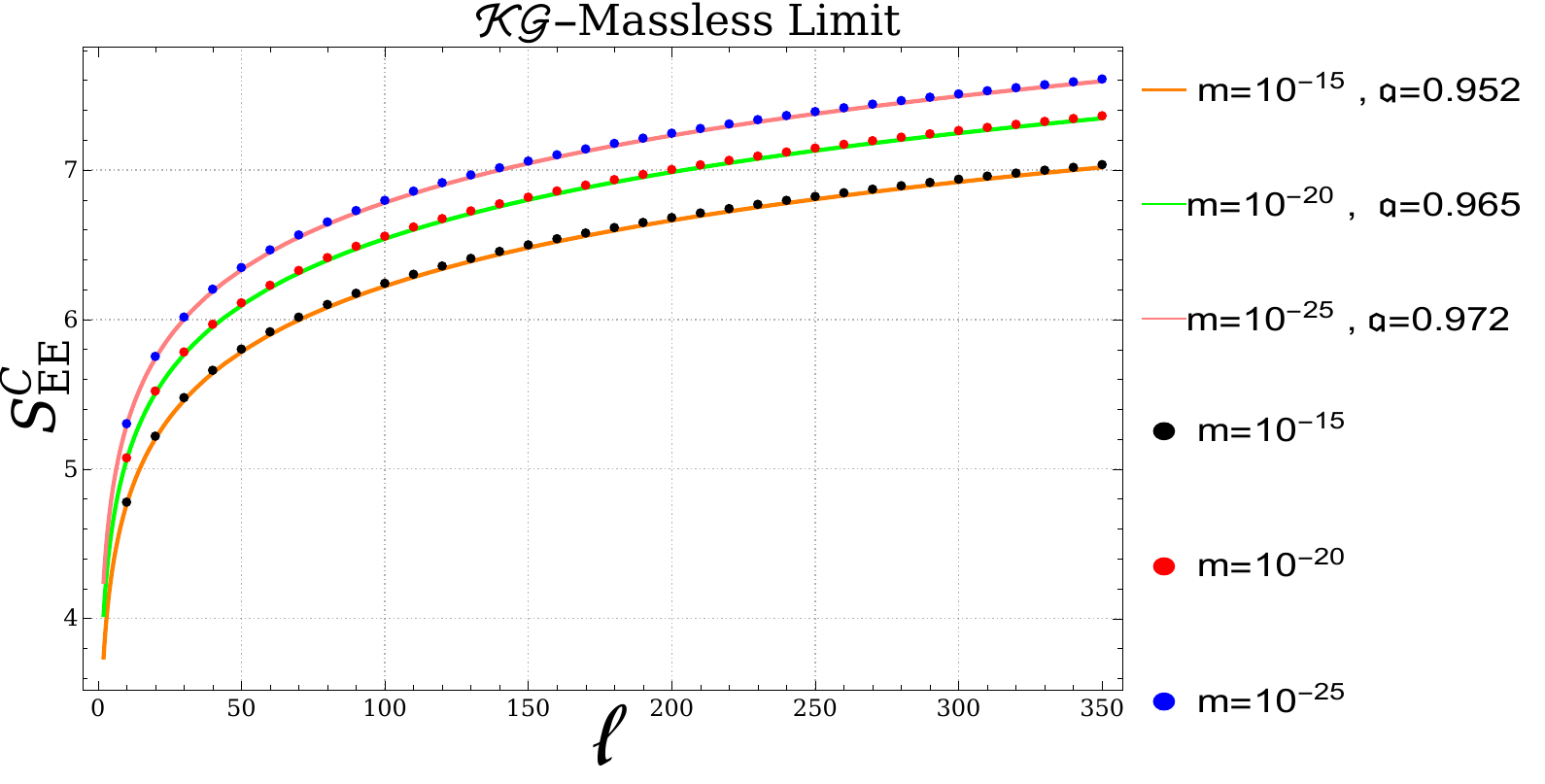}	 
		\caption{ The matrix $\Lambda_\mathcal{A}$ illustrates the three largest eigenvalues as a function of $\ell$ for $m=10^{-20}$ in the left panel. The numerical values and those predicted by equation \eqref{Eq:Eig-masless} for the eigenvalues in the massless case are depicted by points and lines, respectively. In the right panel, $S_{EE}$ is shown as a function of $\ell$ using two methods: numerical analysis (points) and direct calculation from equation \eqref{Eq:EEprediction-massless} (solid lines), with $m$ set to $10^{-15}$, $10^{-20}$, and $10^{-25}$. These figures confirm the agreement between the numerical and theoretical calculations. }  \label{Fig:massless-KG}
	\end{figure}

\begin{figure}[h]
		\centering
		\includegraphics[scale=.39]{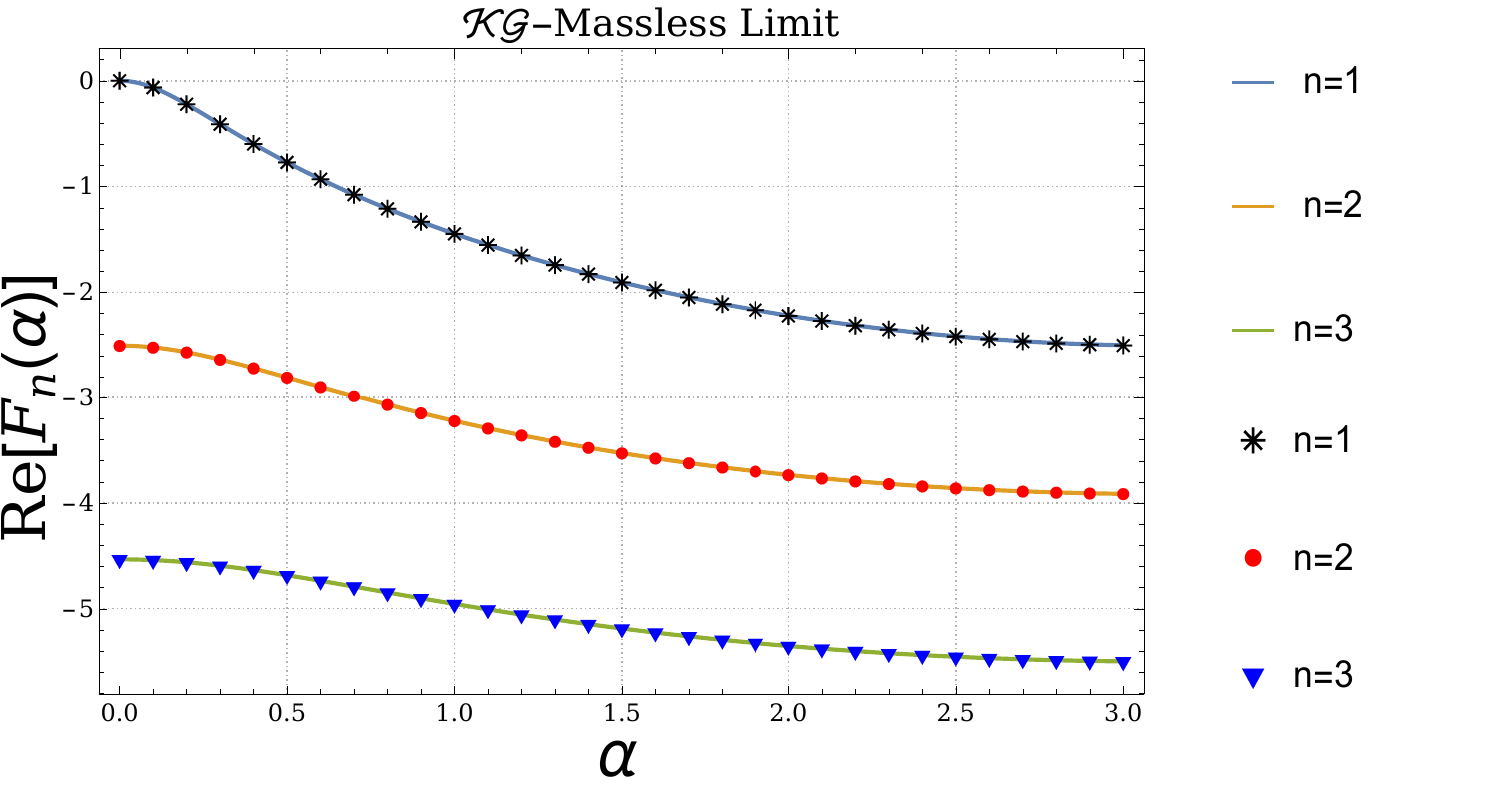}\includegraphics[scale=.39]{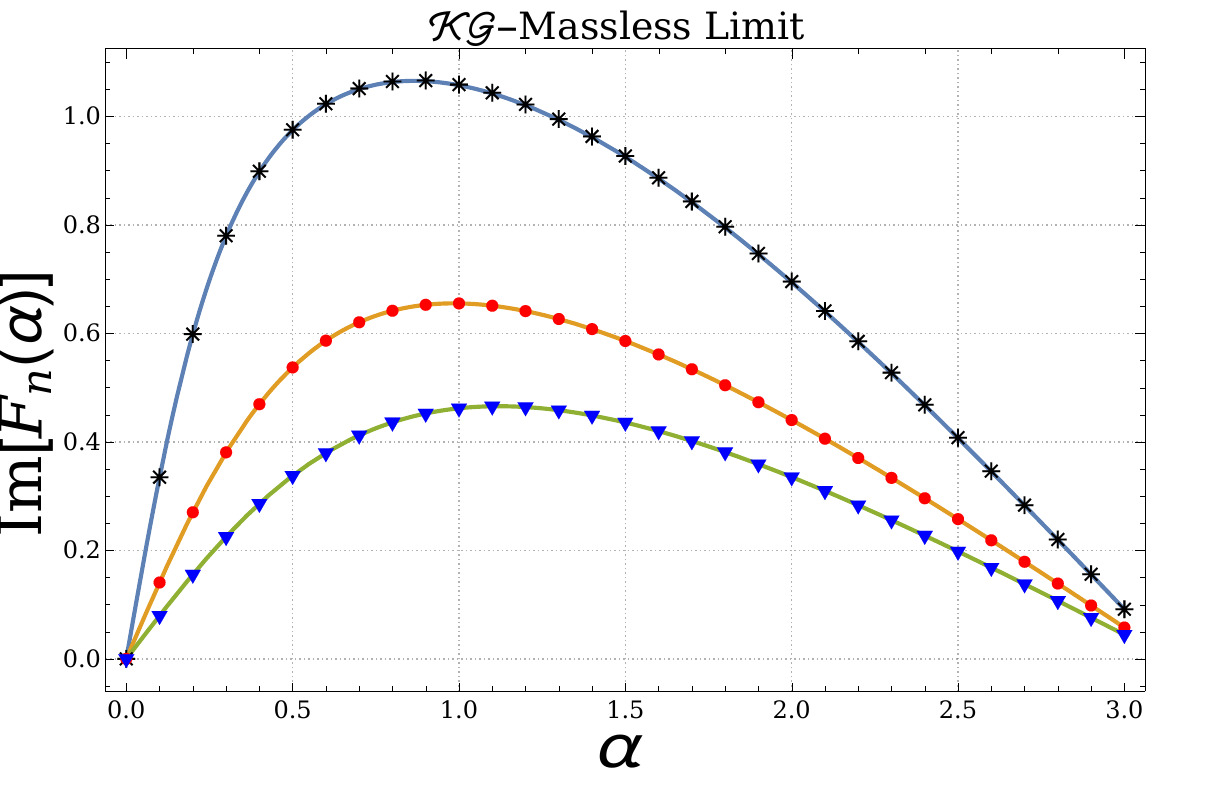} 	
		\includegraphics[scale=.39]{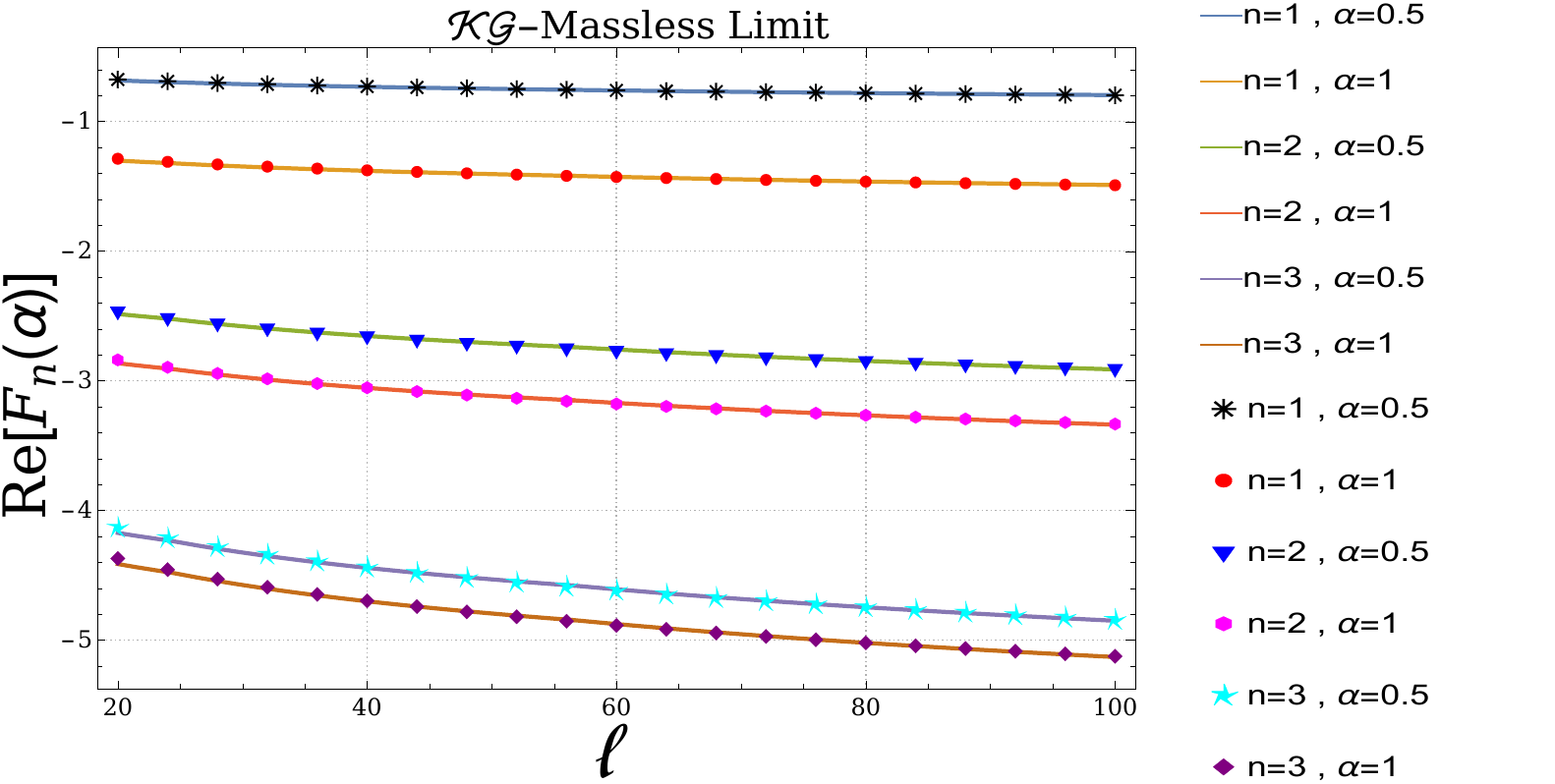}\includegraphics[scale=.39]{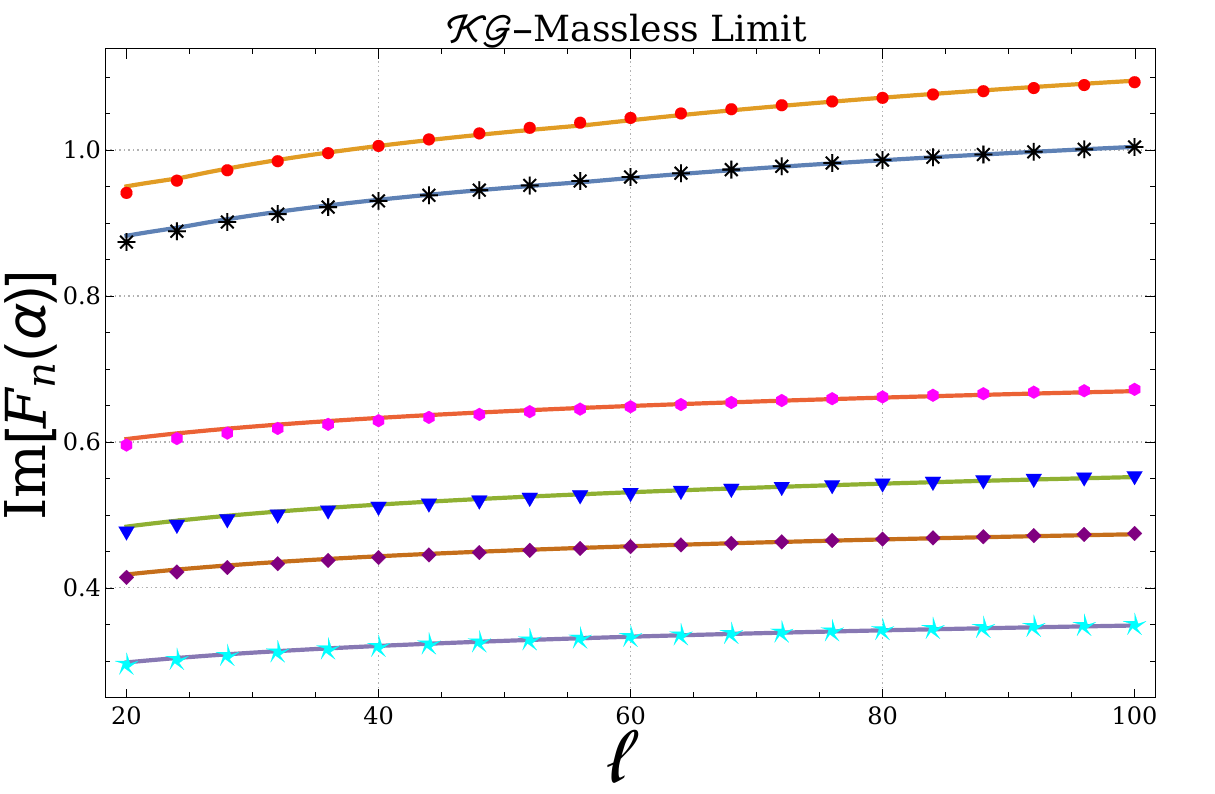} 
		\caption{The comparison between the results of numerical analysis (represented by symbols) and the values predicted by equation \eqref{Eq:Fn-massless} (depicted by the solid line) is illustrated in the graphs showing the real and imaginary components of the function  $F_n(\alpha)$  in the massless limit scenario where we have set $m= 10^{-20}$.  The upper graphs display the behavior with respect to $\alpha$ and $n=1,2$ and $3$ at $\ell=70$. Conversely, the lower graphs illustrate the behavior with respect to $\ell$ for $\alpha =0.5$ and $1$. }  \label{Fig:FnalphanotcnstKG-massless}
	\end{figure}

\begin{figure}[h]
		\centering
		\includegraphics[scale=.39]{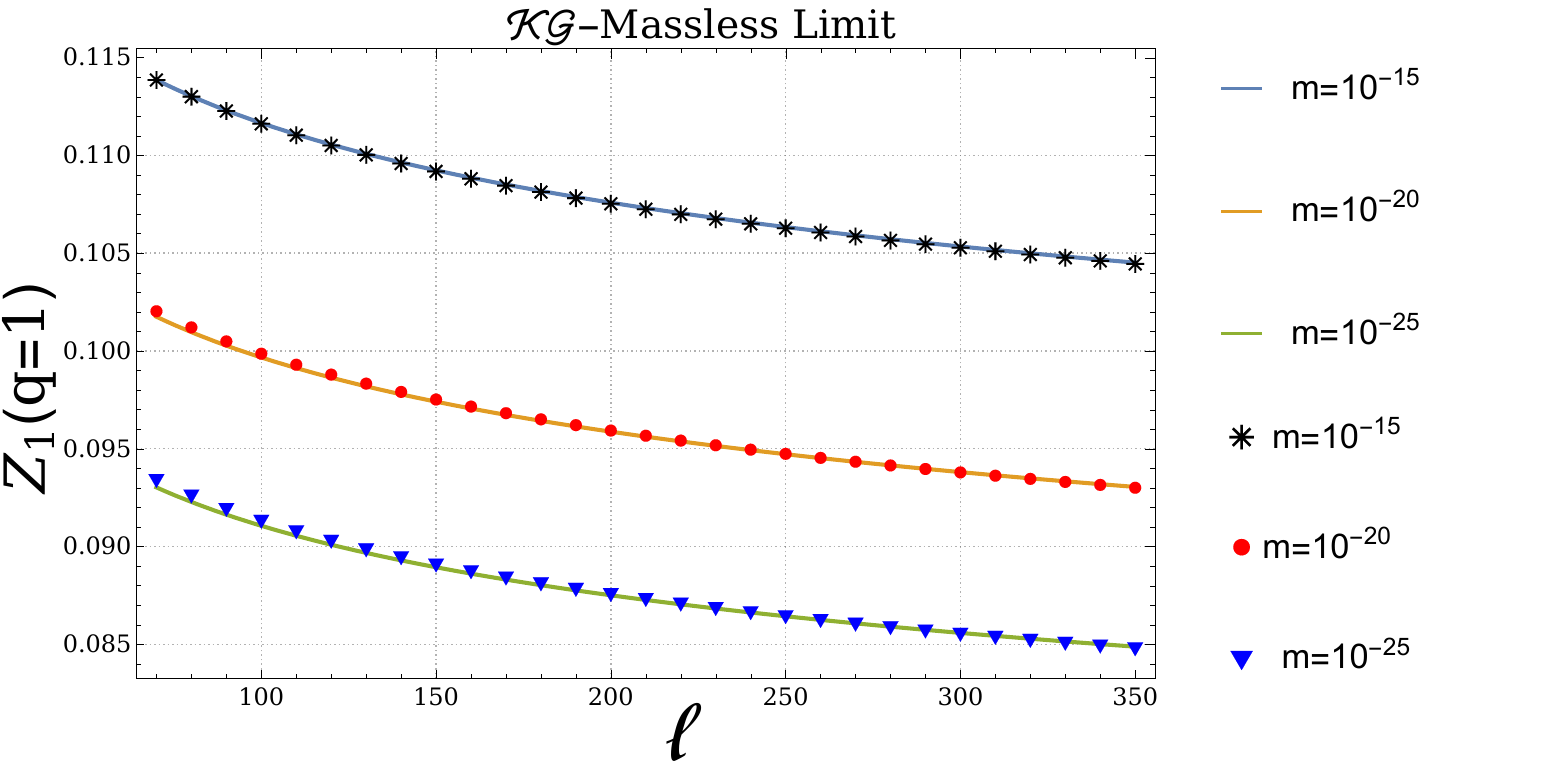}\includegraphics[scale=.39]{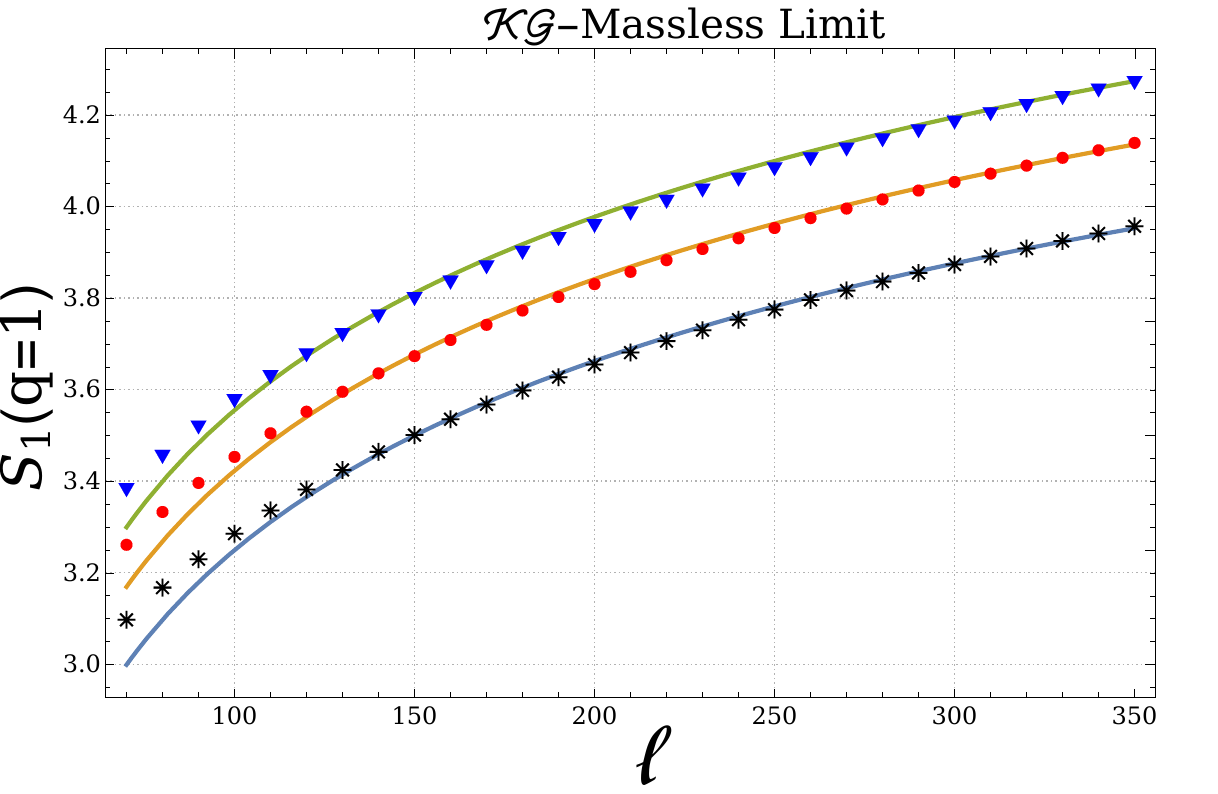} 	
		\includegraphics[scale=.39]{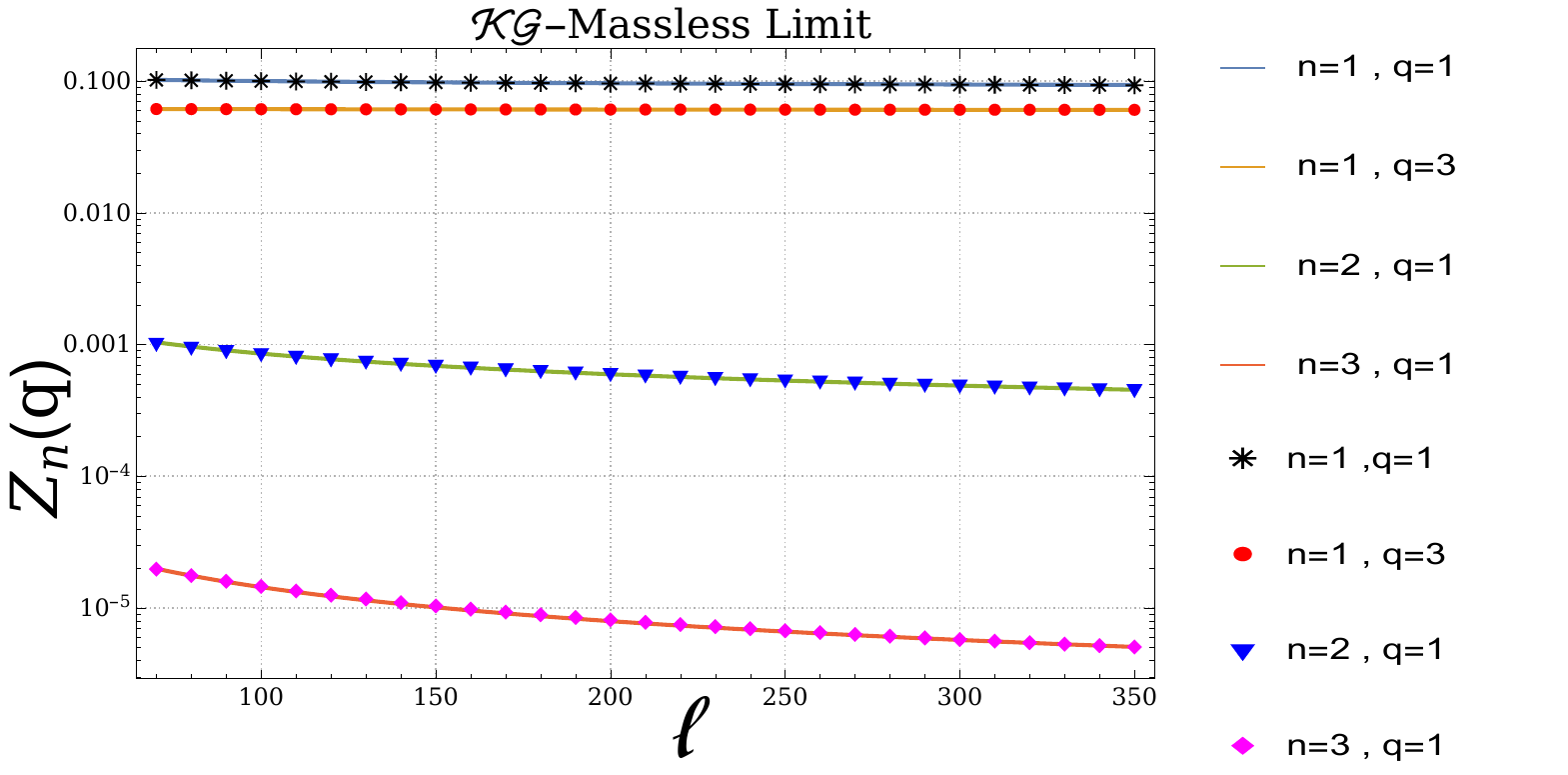}\includegraphics[scale=.39]{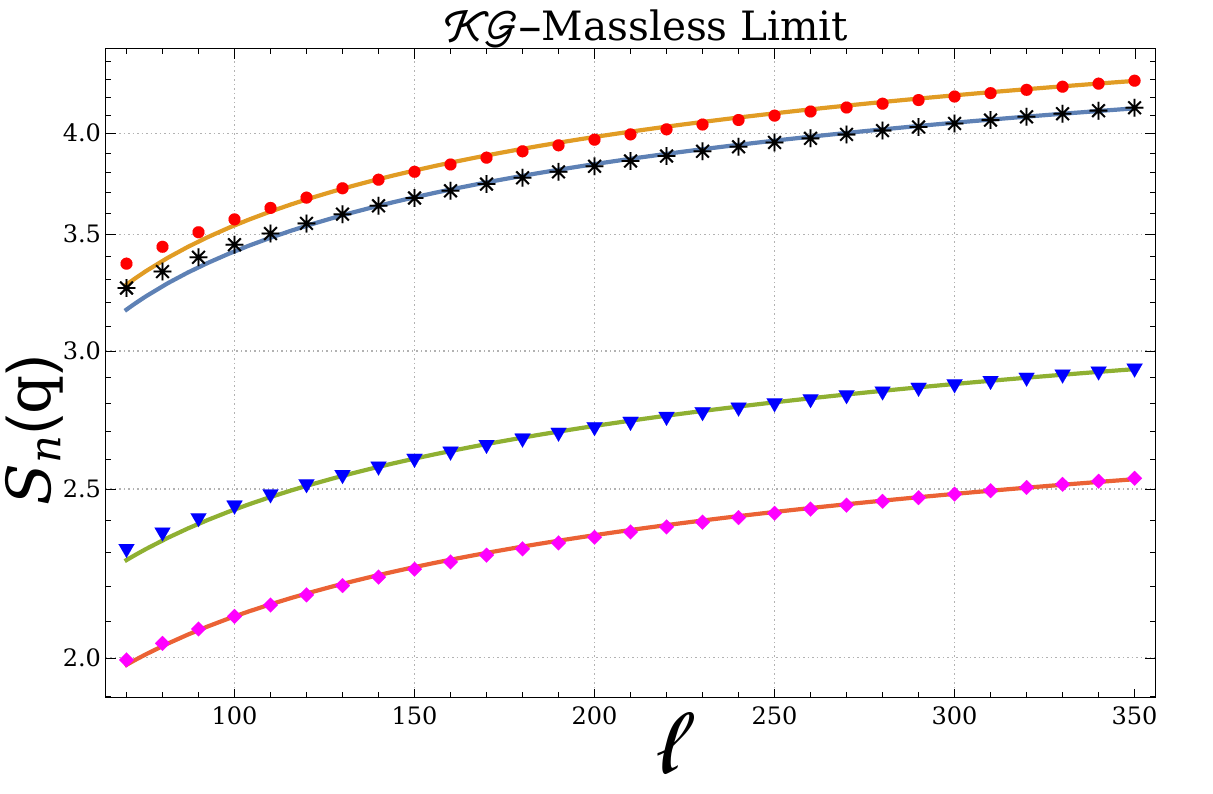} 
		\caption{The values of $\mathcal{Z}_n(q)$ and $ S_n(q)$ in conformal limit, including the numerical results (symbols) against to what obtained from the equation \eqref{Eq:Znalpha-massless} (solid lines) in terms of $\ell$ for $n=1$, $q=1$ and different masses (upper plots) and $m=10^{-20}$, $q=1,3$ and $n=1, 2$ and $3$ (bottom panels) are shown. } 
		\label{Fig:ZnqSnqtKG-massless}
	\end{figure}

	The universal behavior at one-dimensional conformal quantum critical points is a significant outcome of EE, known to be determined by the central charge of the underlying CFT. At the quantum critical point, the EE and the entanglement R\'enyi entropies yield well-known following results \cite{Calabrese:2004eu ,cw,Calabrese:2009qy} 
	\begin{equation}
		S_{EE}^{CFT}=\frac{c}{ 3 }\log\frac{\ell}{ \epsilon }+cnst ,\hspace{4mm}  S_n^{CFT}=\frac{c}{ 6 } \frac{n+1}{n }\log\frac{\ell}{ \epsilon }+cnst.
	\end{equation}
The conformal limit for scalar and fermionic quantum fields has been studied in \cite{Casini:2005zv,Casini:2005rm}. Additionally, the analytical calculations for charged moments of Dirac and complex scalar fields can be found in \cite{Murciano:2020vgh}. In this study, we aim to explore and analyze the behavior governing charged moments of complex scalar fields, and subsequently investigate $\mathcal{Z}_n(q)$ and the symmetry-resolved entropies through numerical calculations based on the eigenvalues of the matrix $\Lambda_\mathcal{A}$. For free complex scalar fields, considering the 
	\eqref{Eq:towpointPKG}, \eqref{Eq:towpointQKG} and after making use of \eqref{Eq:anaPQKG} in the massless limit\footnote{Note by massive and massless limit we mean $m\ell\gg1$ and $m\ell\ll1, $  respectively.}, we obtain 
	\begin{eqnarray}
		\mathbf{P}_{rs}&\simeq&\frac{-4 }{\pi\Big(4( r-s)^2-1\Big) }\\
		 \mathbf{X}_{rs}&\simeq&
	\begin{cases}
		\frac{1}{\pi}\Big(3\log 2-\log m\Big)        & \quad \text{for } \hspace{3mm} r= s\\
		\frac{1}{\pi}\Big(3\log2-\log((r-s)m)-2\Big)        & \quad \text{for} \hspace{4mm} r\neq s\\
	\end{cases}
		\end{eqnarray}
We would like to highlight that the lattice spacing has been fixed at one, resulting in the dimensions of $m\epsilon$ and $\frac{\ell}{\epsilon}$ scaling as $m$ and $\ell$, respectively. Upon ordering the eigenvalues in descending order, our numerical computations reveal that only a small subset of eigenvalues (typically 3 to 4) emerge as significant and hold crucial importance in the calculations. Hence, through interpolating the data points, we can approximate the behavior of eigenvalues of the $\Lambda_{\mathcal{A}}$ matrix as 
\begin{equation}\label{Eq:Eig-masless}
		 \nu_i \sim c_i \Big(3\log2+\log\frac{1}{m} \, (\frac{1}{3}\log\ell+1)\Big)^{\frac{1}{2}}	,\hspace{7mm}  i=1,2,3,4
	\end{equation}
	 $c_i$'s are numerical coefficients. In order to do the numerical test, the three dominant eigenvalues of the matrix $\Lambda_\mathcal{A}$ have been graphed on the right side of Figure \ref{Fig:massless-KG}. The comparison between the numerical analysis and equation \eqref{Eq:Eig-masless} reveals a high level of consistency in the obtained results. The infrared divergence at $m=0$ is a result of the well-known zero mode, as mentioned in \cite{AparicioAlcalde:2007td,Bianchini:2016mra}. When considering the EE, utilizing the equation \eqref{Eq:SEE} and the eigenvalues expression \eqref{Eq:Eig-masless} yields
	\begin{equation}\label{Eq:EEprediction-massless1}
		S_{EE}(\ell)\sim \frac{\mathfrak{a}}{3}\log\ell+cnst,
	\end{equation}
In numerical analysis, it is found that $\mathfrak{a}$ is approximately equal to 1. However, when examining complex scalar fields under the conformal limit, a distinct outcome emerges
	 	\begin{equation}\label{Eq:EEprediction-massless}
	 S^C_{EE}(\ell)\sim \frac{2}{3}\log\ell+cnst,
	 \end{equation}
note that the 2 factor is introduced for complex fields. We would like to emphasize that the resulting EE at the leading order aligns closely and coherently with past computations in this context. For instance, the entropic $c$-function transforms into $c=\ell\,\frac{d}{d\ell}S_{EE}=\frac{2}{3}$ \cite{Casini:2005zv}.
In order to compute the SREE, one can write 
	\begin{eqnarray}\label{Eq:Fn-massless}
		F_n(\alpha)&\simeq& \log\Big [\prod_{i=1}^4 \frac{2^{n}}{(\nu_i+1)^{n}-e^{i\alpha}(\nu_i-1)^{n} }\Big]\\
	\label{Eq:Znalpha-massless}
		{Z}_n(\alpha)&\simeq&\prod_{i=1}^4\frac{2^{2n}}{(\nu_i-1)^{2n}+(\nu_i+1)^{2n}-2(\nu_i^{2}-1)^{n}\cos \alpha }, 
	\end{eqnarray}
where $\nu_i$'s are given by \eqref{Eq:Eig-masless}, therefore, by using \eqref{Eq:defZnq} and \eqref{Eq:SReen},   $\mathcal{Z}_n(q)$ and the symmetry-resolved entropies can be obtained. In Figure \ref{Fig:FnalphanotcnstKG-massless},  we have plotted the $	F_n(\alpha)$ function for $n=1,\,2$ and $3$, and different certain values of $\alpha$  
 and check the graphs against the numerical ones. In Figure \ref{Fig:ZnqSnqtKG-massless}, the plots of $\mathcal{Z}_n(q)$ and $ S_n(q)$ are displayed, utilizing equation \eqref{Eq:Znalpha-massless} (solid lines) with respect to $\ell$ for $n=1$, $q=1$ (top graphs) and $q=1,3$ and $n=1, 2, 3$ (bottom graphs).  According to the graphs, it would be satisfactory to only take into account the four largest eigenvalues in the numerical analysis.

\begin{figure}[h]
		\centering
		\includegraphics[scale=.34]{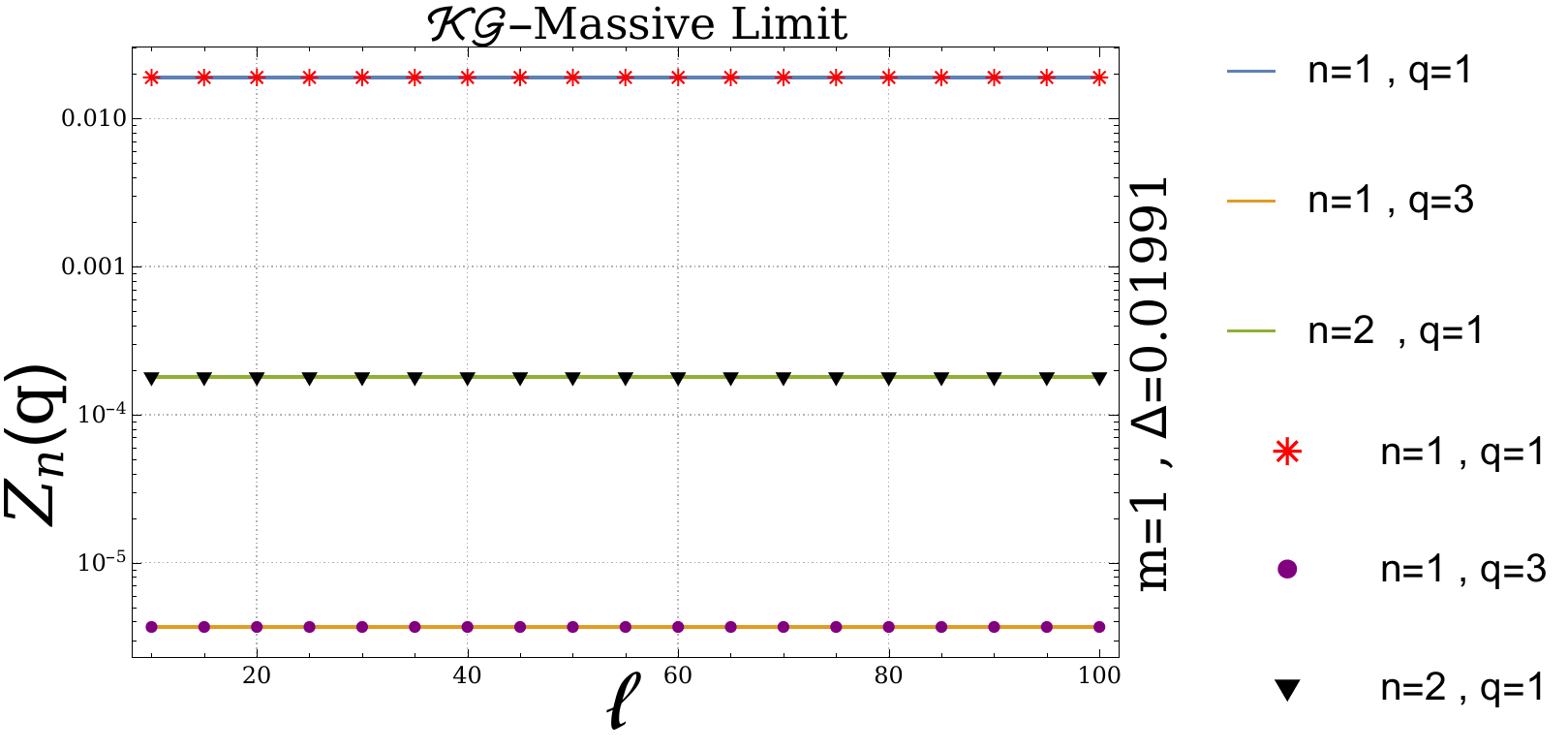} \includegraphics[scale=.34]{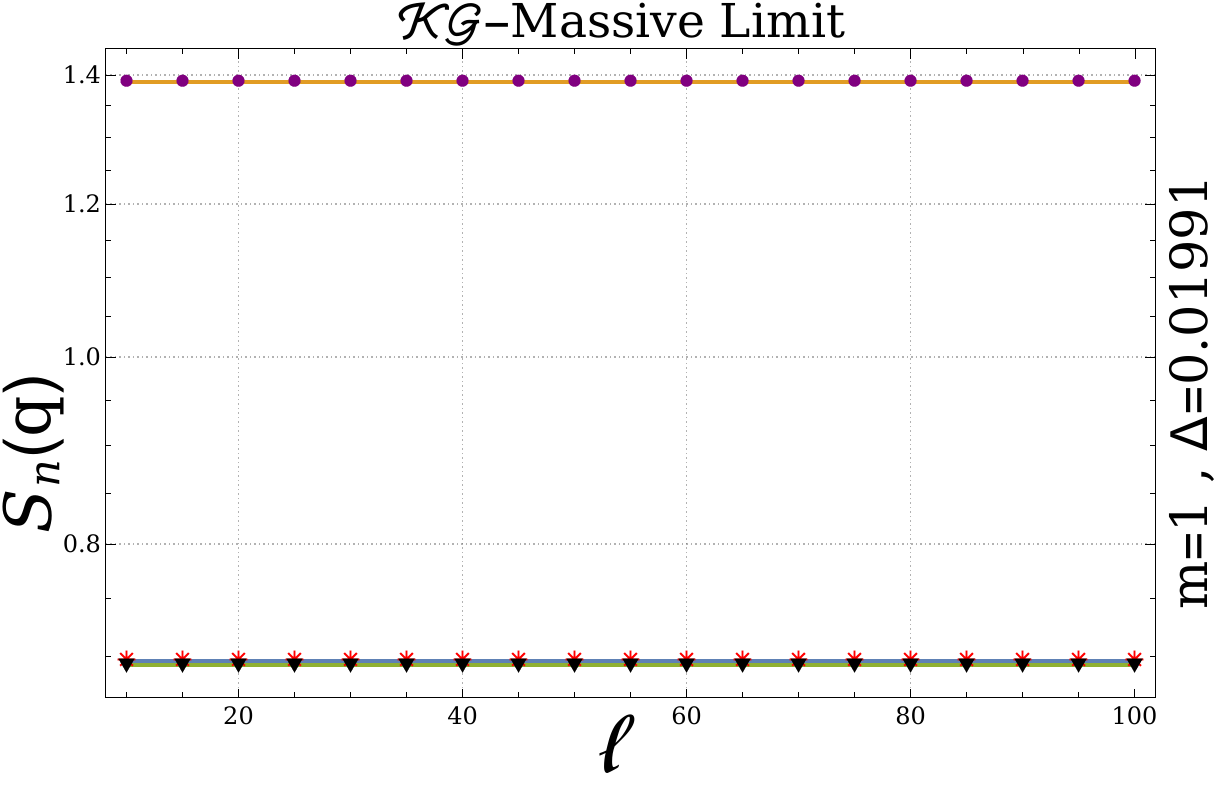}	 
		\caption{In the massive limit, the values of $\mathcal{Z}_n(q)$ and $S_n(q)$ are presented, alongside numerical results (depicted by symbols) compared to those obtained from equation \eqref{Eq:Zn1alpha-m} (illustrated by solid lines) as functions of $\ell$ for $n=1,\,2$ and $q=1$, as well as for $n=1,\, q=3$ when $m=1$. The results demonstrate consistency.  } \label{Fig:massiveZnSn-KG}
	\end{figure}

	In the case of a  massive limit, it is demonstrated that only two eigenvalues are crucial in the computations. When the subsystem's length is sufficiently large, the eigenvalues become equal and are not affected by the length of the subsystem. It is then proven that they align closely with the following expression
	\begin{equation}\label{Eq:Eigmassive}
		\nu \simeq1+\Delta(m) ,
	\end{equation}
where the parameter $\Delta\ll 1$ is a real positive value that relies on the mass. The EE can be acquired in this case as 
	\begin{equation}\label{Eq:EEprediction-massive-complex}
		S_{EE}^{ m}\sim \Delta\big[1+\log(\frac{2}{\Delta})\big].
	\end{equation}
In order to find SREE, it is important to observe that the expression $Z_n^m(\alpha)$ can be found in \eqref{Eq:Znalpha}, with our focus being on the two highest eigenvalues,  
	\begin{figure}[h]
		\centering
		\includegraphics[scale=.34]{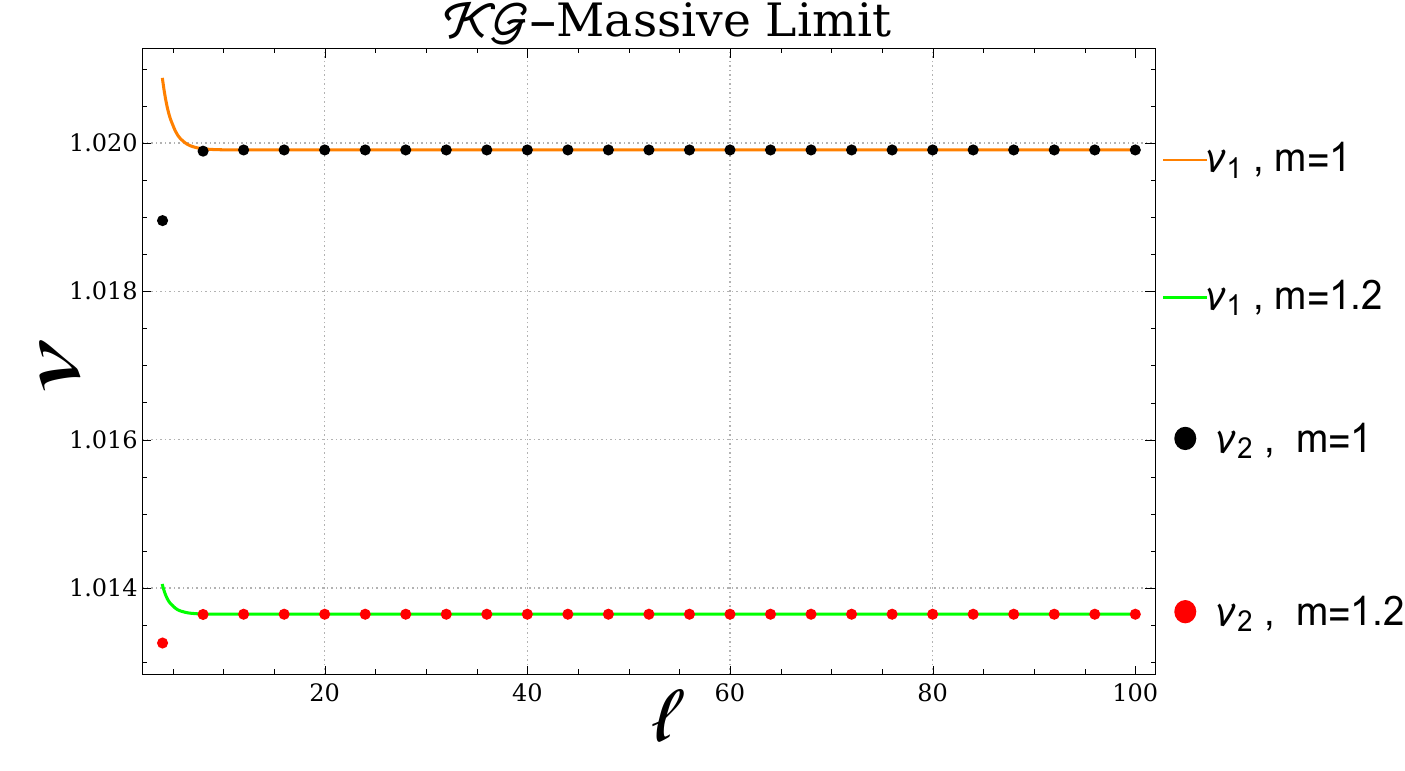} \includegraphics[scale=.33]{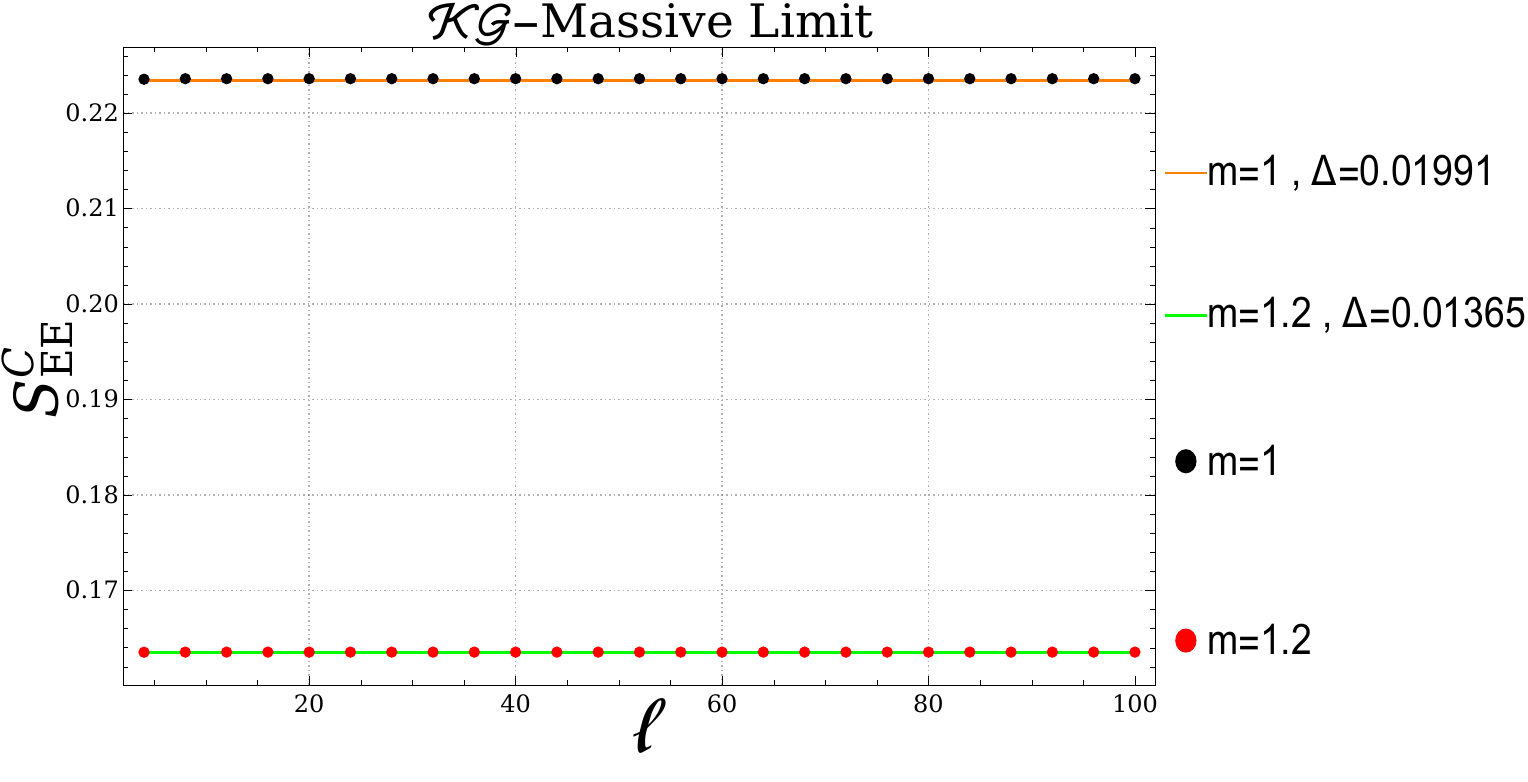}	 
		\caption{ In the left panel, the top two eigenvalues of the matrix $\Lambda_\mathcal{A}$ in the massive limit are displayed as a function of $\ell$ for $m=1$ and $1.2$. On the right panel, numerical results (points) and the corresponding results from equation \eqref{Eq:EEprediction-massive-complex} (solid lines) are shown in relation to $\ell$ for $m=1$ and $1.2$. The strong agreement between the numerical data and the predicted values is clearly visible.} \label{Fig:massive-KG}
	\end{figure}
at the leading order around $\nu=1$ one obtains	
	\begin{equation}\label{Eq:Zn1alpha-m}
		{Z}_n^{m}(\alpha)\sim\frac{16^n}{\big[-2^n \cos\alpha (n \Delta +2) \Delta ^n+\Delta ^{2 n}+4^n n \Delta +4^n\big]^2}\,.
	\end{equation}
The graph in Figure \ref{Fig:massive-KG} illustrates the top two eigenvalues of the matrix $\Lambda_\mathcal{A}$ with respect to $\ell$ for two distinct mass values. Additionally, it displays the numerical plots of the EE compared to the values predicted by equation \eqref{Eq:EEprediction-massive-complex} in terms of $\ell$, which are in agreement. In addition, within Figure \ref{Fig:massiveZnSn-KG}, the outcomes of $\mathcal{Z}^m_n(q)$ and $ S^m_n(q)$ calculated from equation \eqref{Eq:Zn1alpha-m} are juxtaposed with the numerical findings, demonstrating identical results.\\   In conclusion, we employed the estimated eigenvalues for the correlation matrix, simplifying the numerical computations compared to traditional methods. The outcomes for the SREE are well-aligned with prior research in the field \cite{Murciano:2019wdl , Murciano:2020vgh}. 

	\section{Non-local quantum fields}
		 \label {none-local}
In local quantum field theories and for a gapped Hamiltonian with an ultraviolet fixed point, it has been demonstrated that due to the short-ranged interactions, the EE follows the area law \cite{Eisert:2008ur}.\footnote{Even though one-dimensional critical systems go against the usual pattern by having the EE increase logarithmically with the subsystem size \cite{Calabrese:2009qy,Casini:2009sr}.}  However, it is observed that by considering the non-local random interactions and non-local field theories, the EE follows the volume law
\cite{Vitagliano:2010db}.
	For a simple class of the non-local scalar field theories, let us assume the following Hamiltonian
	\begin{equation}
	{\cal H}=\tfrac{1}{2}\int dx   \left[ (d\phi/dt)^2 + C \phi e^{A (-\partial^2)^{w/2} } \phi \right],
	\label{Hamiltonian:NL}
	\end{equation}
	where positive constants $ A$ and $ w$ are used to determine the magnitude of non-locality, and for simplicity, we set $ C=1 $ by re-scaling the time. In the reference \cite{Shiba:2013jja}, it has been demonstrated that the ground state EE of this particular non-local field follows the volume law if the subsystem size $\ell$ is less than $A$. Moreover, for $w=1$ and $w=2$, the theory offers an analytical estimate that supports the numerical analysis. In this section concerning non-local field theories of this kind for $A\gg\ell$, we provide an evaluation for the SREE through numerical computations. The comparison against numerical analysis is carried out for two values of $w=1\,,2$. Additionally, we demonstrate formulas for the symmetry-resolved R\'enyi entropies, proving their $q$ and $n$ independence. Consequently, by calculating the EE, we derive analytical expressions for the configurational and number entropies as defined in \eqref{Eq:SvN}.\\	
	 In order to compute the EE, it is technically satisfactory to acquire the correlation function representation for this non-local theory. Consequently, $ \mathbf{P}_{{\mathcal{NL}}}$ and $ \mathbf{X}_{{\mathcal{NL}}} $ turn to
	\begin{equation}
	(\mathbf{P}^w_{{\mathcal{NL}}})_{rs}=\int_{-\pi}^{\pi} \dfrac{dp}{2\pi} e^{i p(r-s)}
	\exp [A/2(2-2\cos p )^{w/2} ]
	\label{EQ:PPNL}
	\end{equation}
	\begin{equation}
	(\mathbf{X}^w_{{\mathcal{NL}}})_{rs}=\int_{-\pi}^{\pi} \dfrac{dp}{2\pi} e^{i p(r-s)}
	\exp [-A/2 (2-2\cos p )^{w/2} ]
	\label{EQ:XXNL}
	\end{equation} 
	The solution to the integrals mentioned above can be expressed using a combination of the generalized hypergeometric function $\, _\mathfrak{p}F_\mathfrak{q}(\mathfrak{a}_1,...,\mathfrak{a}_\mathfrak{p};\mathfrak{b}_1,...,\mathfrak{b}_\mathfrak{q};z)$. As an illustration, for $(\mathbf{P}^{w_{even}}_{{\mathcal{NL}}})_{rs}$, they will take on a general form as follows	\begin{multline}\label{hyper}
	\, _\frac{w}{2}F_\frac{w}{2}\left(\frac{1}{w},\frac{3}{w},...,\frac{w-1}{w};\frac{2}{w},\frac{4}{w},...,1;2^{w-1} A\right)-( r-s)^2\, _\frac{w}{2}F_\frac{w}{2}\left(\frac{3}{w},\frac{5}{w},...,\frac{w+1}{w};\frac{4}{w},...,\frac{w+2}{w};2^{w-1} A\right)
	\\
	+\cdots+(-1)^{i+1}\mathcal{C}_i\, _\frac{w}{2}F_\frac{w}{2}\left(\frac{2i-1}{w},\frac{2i+1}{w},...,\frac{2(i+\frac{w}{2})-3}{w};\frac{2i}{w},\frac{2(i+1)}{w},...,\frac{2(i+\frac{w}{2})-2}{w};2^{w-1} A\right)+\cdots
	\end{multline} 
	where  $i=1,2,...,Max(| r-s|)+1$ and $\mathcal{C}_i$'s are integer coefficients.  In this case, one should also find the eigenvalues for the matrix  $\Lambda_\mathcal{A}= \sqrt{\mathbf{X}_\mathcal{A} \mathbf{P}_\mathcal{A}}$. In the following, we investigate this entropy equation in detail. In order to generalize calculations to the non-local scalar field, one must put $\omega^2(p)=e^{A p^{\omega }}$ in \eqref{Eq:Phi}. In various specific instances, the generalized hypergeometric function is automatically transformed into other distinct functions. For the sake of simplicity, this study primarily concentrates on the scenarios where $ w  = 1$ and $2$ in \eqref{hyper}. Thus, one obtains
	\begin{eqnarray}
	(\mathbf{P}^{w=1}_{\mathcal{NL}})_{rs}&=&J_{2( r-s) }(i A)+i\pmb{E}_{2( r-s) }(i A),
	\label{Eq:weqal1}\\	
	(\mathbf{P}^{w=2}_{\mathcal{NL}})_{rs}&=&(-1)^{(r-s)}e^A I_{(r-s)}(A), 
	\label{Eq:weqal2}
	\end{eqnarray}
where $J_{2 r }$ and $\pmb{E}_{2 r }$ are Bessel functions of the first kind and Weber functions, respectively, and $I_{r}$ is the modified Bessel function of the first kind. It should be noted that to calculate $\mathbf{X}_{\mathcal{NL}} $, it is enough to replace $-A$ with $A$.

\subsection{Numerical analysis for SREE }
\begin{figure}[h]
	\centering
	\includegraphics[scale=.27]{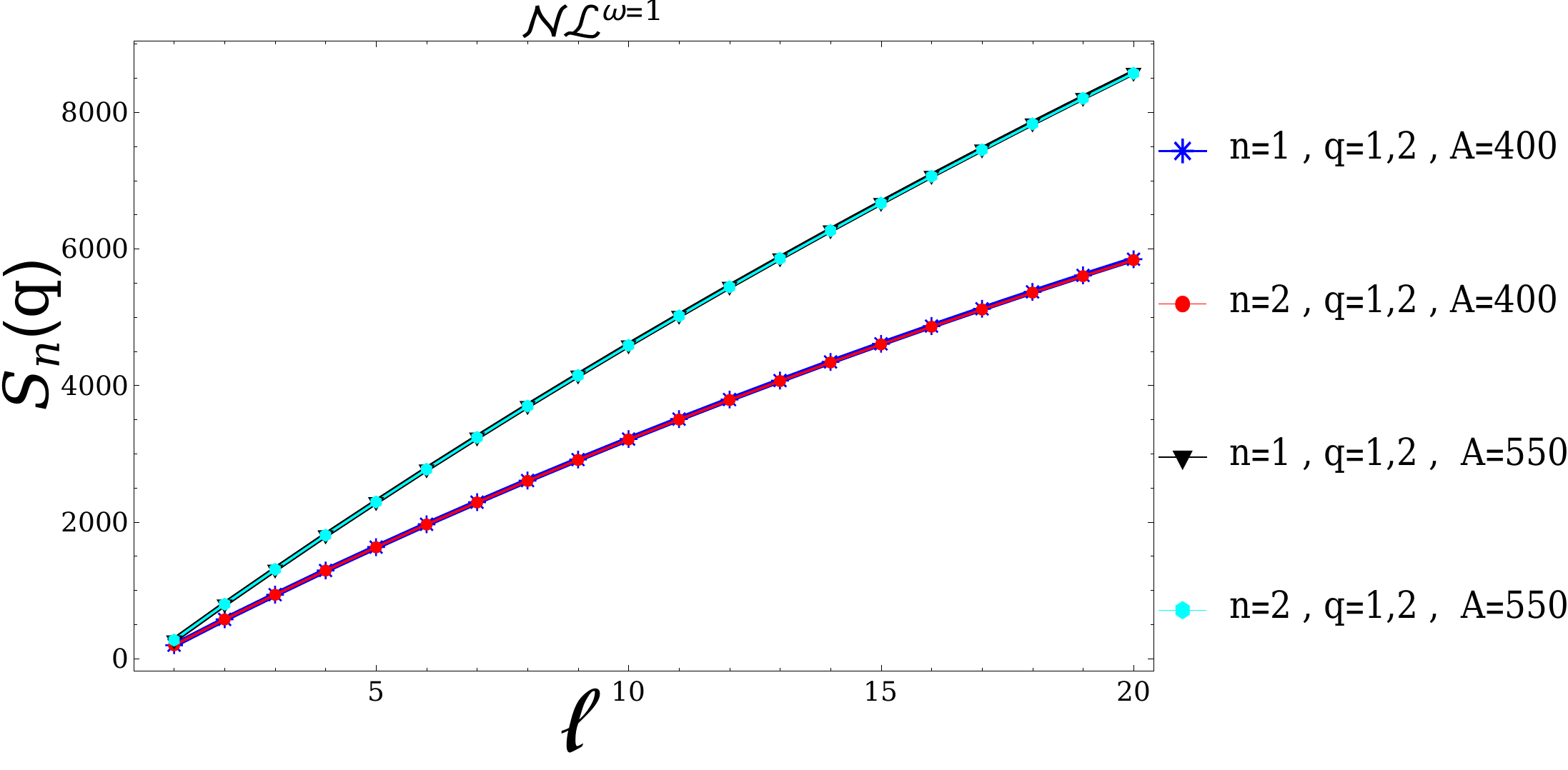}\includegraphics[scale=.27]{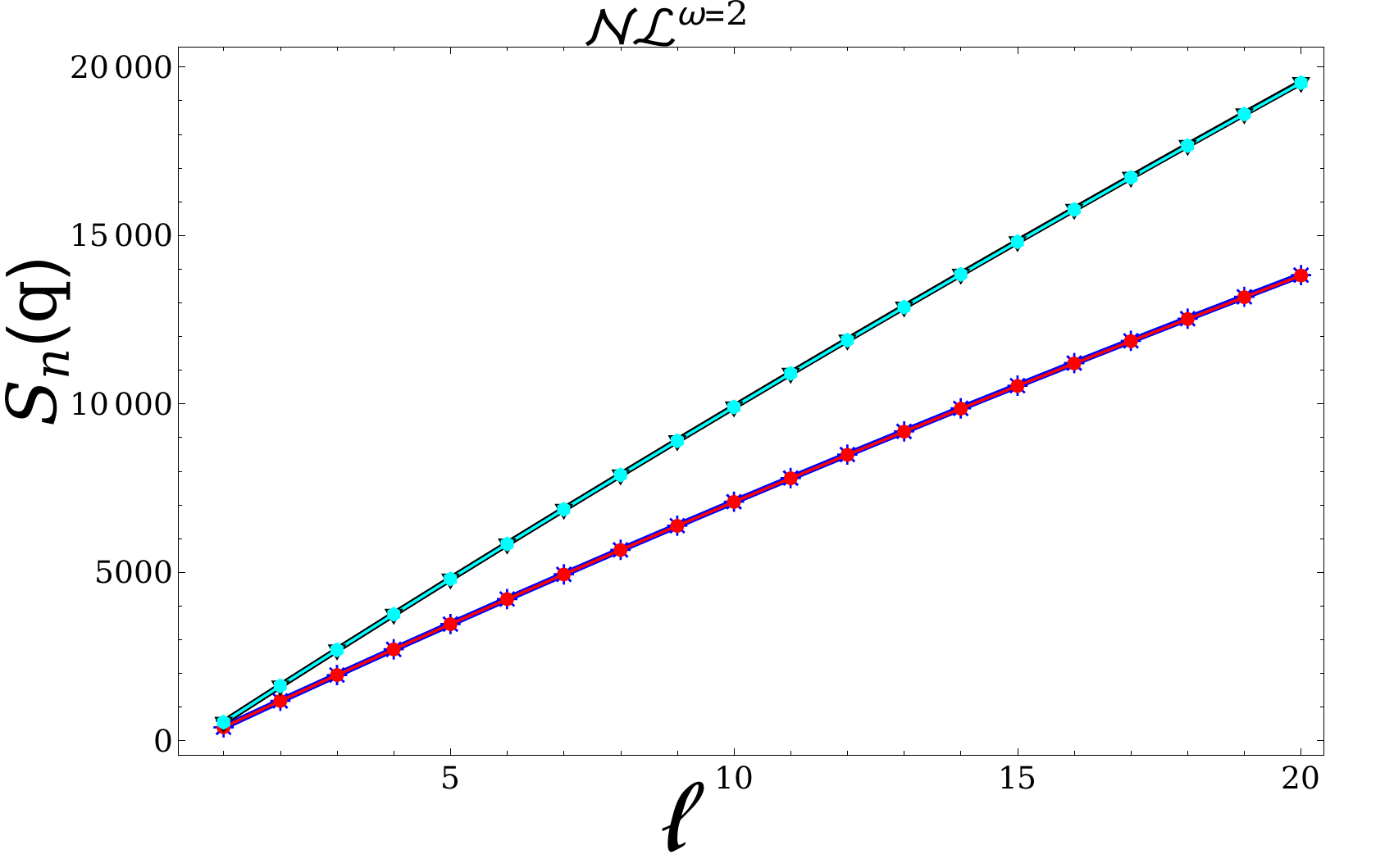} 	 
	\caption{ The numerical results for $S_n(q)$ are plotted for
		$w=1$ (left) and $w=2$ (right). $w$ and $A$ are the parameters that increase the amount of non-locality. We set $A = 400,\, 550$ and $ n = 1,\, 2 $ and $ q = 1,\, 2$.  It can be seen that the behavior of
		 $S_n(q)$ is independent of $n$  and $q$, which confirms equation \eqref{Eq:SnqAlarge1}. } \label{Fig:SNQNLAlarge}
\end{figure}

 In order to analyze the symmetry-resolved moments and the SREE, we will examine the cases where $w=1$ and $w=2$, assuming that the subsystem is significantly smaller than $A$. Under these conditions, the correlation functions can be expanded as described below
\begin{eqnarray}
(\mathbf{P}^{w=1}_{{\mathcal{NL}}})_{rs} &\sim& (-1)^{(r-s)} e^A \sqrt{ \dfrac{2}{\pi A} }\big [1-\dfrac{1}{A} (2( r-s)^2 -\dfrac{1}{8} )+\dots\big ] \label{Eq:asyw1p}\\
(\mathbf{X}^{w=1}_{{\mathcal{NL}}})_{rs}&\sim& \dfrac{2}{\pi A} [1-\dfrac{1}{A^2} (4 ( r-s)^2 -1)+\dots]  \label{Eq:asyw1x}\\
(\mathbf{P}^{w=2}_{{\mathcal{NL}}})_{rs}&\sim& (-1)^{(r-s)} \dfrac{e^{2A}}{\sqrt{2\pi A}} [1-\dfrac{1}{2A}\Big(( r-s)^2 -\dfrac{1}{4}\Big ) +\dots ]  \label{Eq:asyAl3}\\
(\mathbf{X}^{w=2}_{{\mathcal{NL}}})_{rs} &\sim& \dfrac{1}{\sqrt{2\pi A}} [1-\dfrac{1}{2A}\Big(( r-s)^2 -\dfrac{1}{4}\Big ) +\dots ]
\label{Eq:asyAl4}
\end{eqnarray}
In this case, the eigenvalues of the matrix are  $\nu _i\sim e^{w A/2}$ \cite{Shiba:2013jja}. By making use of \eqref{Eq:Znalpha}, we can write $Z^{A\gg \ell}_n(\alpha)$ as follows
\begin{equation}
Z^{A\gg \ell}_n(\alpha)\simeq\bigg(\frac{2^{2n}}{(e^\frac{A w}{2}-1)^{2n}+(e^\frac{A w}{2}+1)^{2n}-2(e^{Aw}-1)^{n}\cos\alpha }\bigg)^\ell.
\end{equation} 
In the present case, the integral of equation \eqref{Eq:defZnq} can be calculated analytically, and $\mathcal{Z}^{A\gg \ell}_n(q)$ turns to the following form
\begin{equation}
\label{eq:ZnqAlarge}
\mathcal{Z}^{A\gg \ell}_n(q)\simeq \exp\big[- \Big(2 (n-1)\ell+1\Big)\frac{A w}{2}\Big].
\end{equation} 
Finally, using the equation \eqref{Eq:SReen}, one can calculate the symmetry-resolved R\'enyi entropies as
\begin{equation}
\label{Eq:SnqAlarge1}
S^{A\gg \ell}_n(q)= -\frac{1}{2}w\, A+\ell\, w\, A,
\end{equation}
which is independent of $n$. Since the same result holds for SREE, one can conclude that the symmetry-resolved entropies are found to be independent of both $n$ and $q$ in the case of SREE. Additionally, it is noted that $\mathcal{Z}^{A\gg \ell}_n(q)$ and the probability $p(q)$ have consistent values across different sectors. To verify this numerically, Figure \ref{Fig:SNQNLAlarge} illustrates $S^{A\gg \ell}_n(q)$ as a function of $\ell$ for $w = 1$ (left) and $w = 2$ (right) with two distinct values of $A$.  Upon reviewing the computations and numerical analysis, it can be deduced that this model upholds the equipartition of the EE.

	\subsection{Comment on the EE in non-local QFTs}\label{een}
\begin{figure}[h]
		\centering
		\includegraphics[scale=.29]{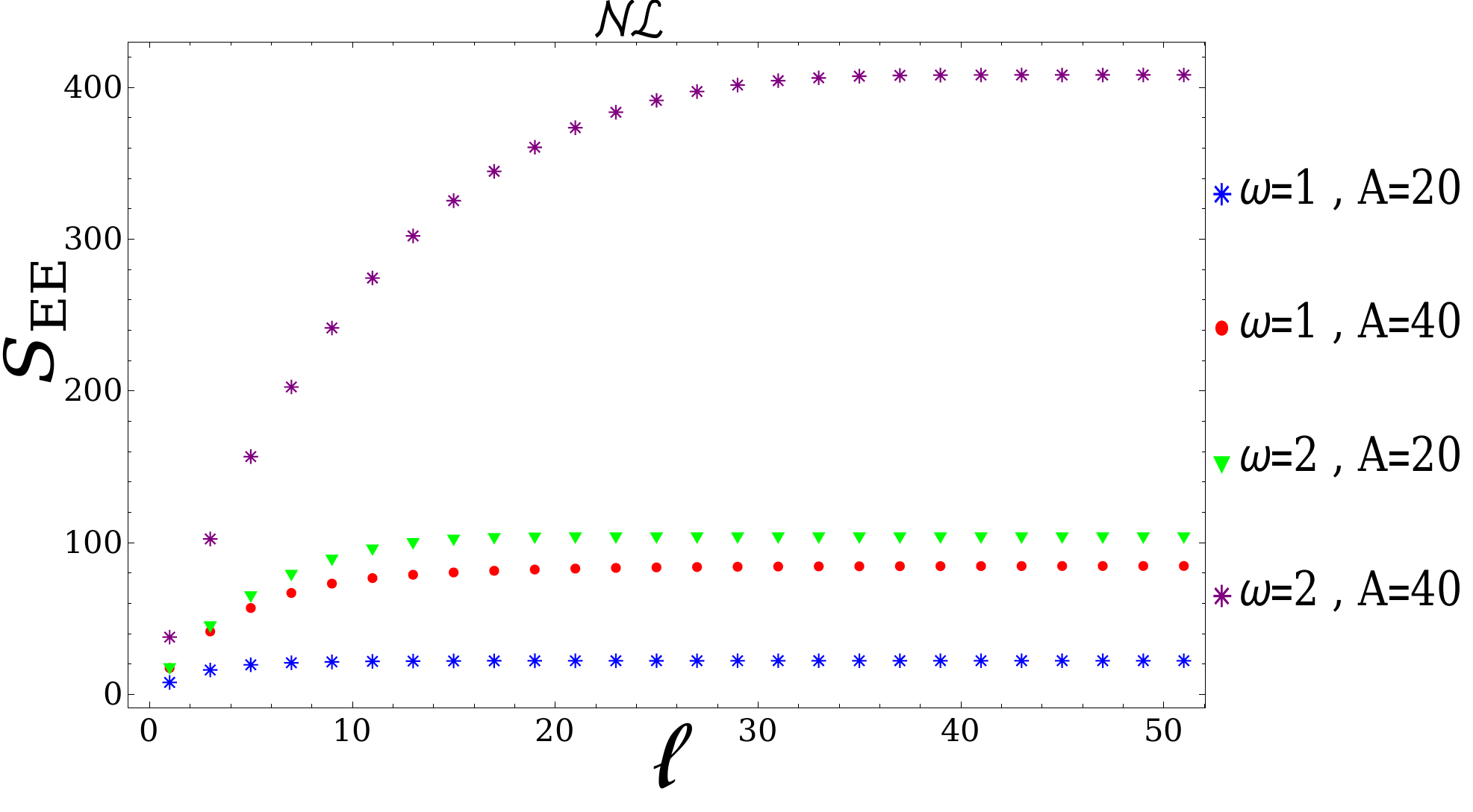}\includegraphics[scale=.27]{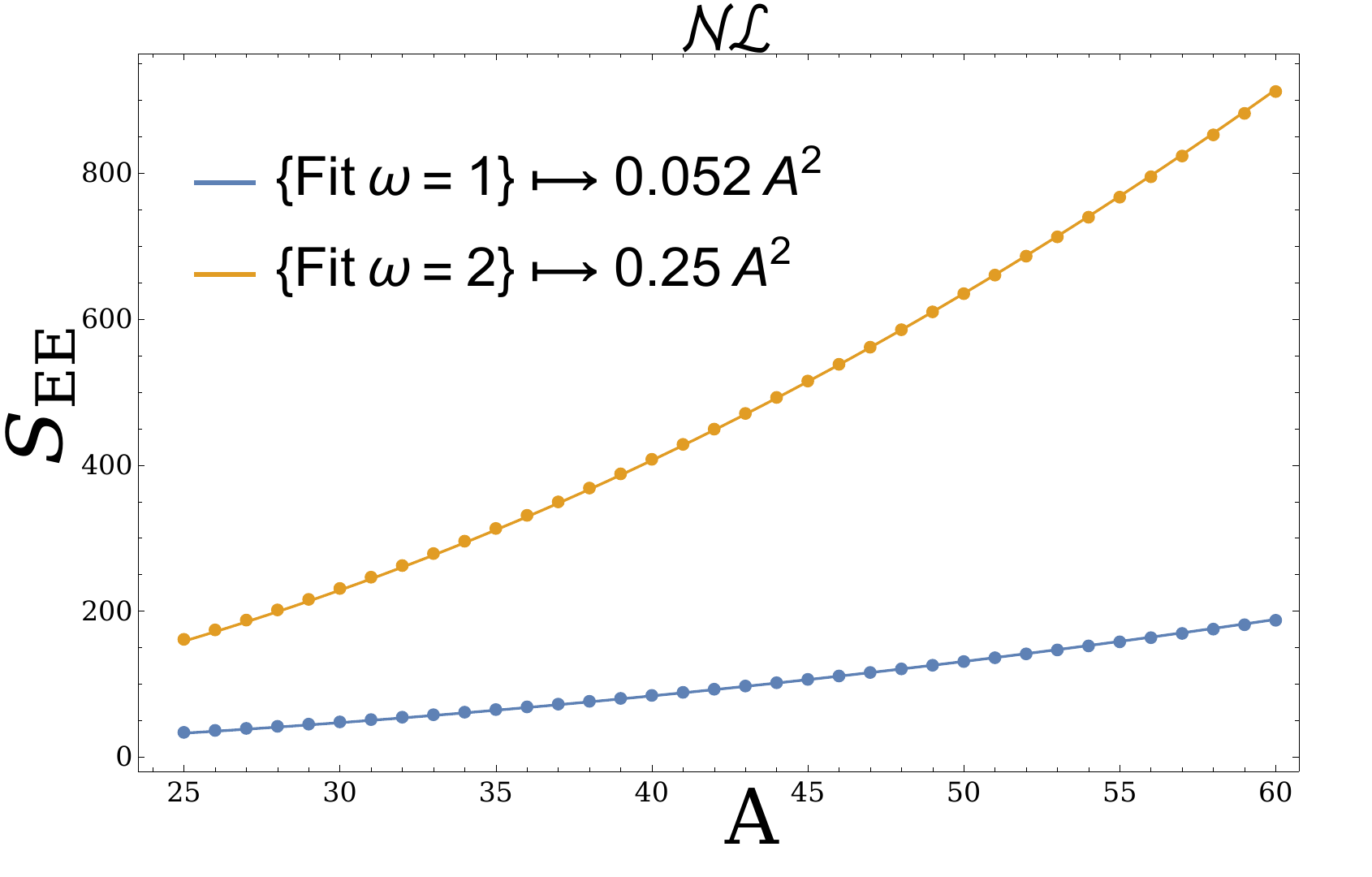} 	 
		\caption{EE is calculated for non-local scalar fields with $ \ell\gg A $. In the left panel, $ S_{EE}$ is graphed based on subsystem length for $w=1 $ and $A=20$ (blue dots), $w=1$ and $A=40$ (red dots), $w=2$ and $A=20$ (green dots), and $w=2$ and $A=40$ (purple dots). The right panel illustrates the relationship with $A$ for $\ell= 85$. Through fitting, it is evident that $ S_{EE}$ scales proportionally to $A^2$. }\label{Fig:SeeaNL}
	\end{figure}

\begin{figure}[h]
		\centering
		\includegraphics[scale=.28]{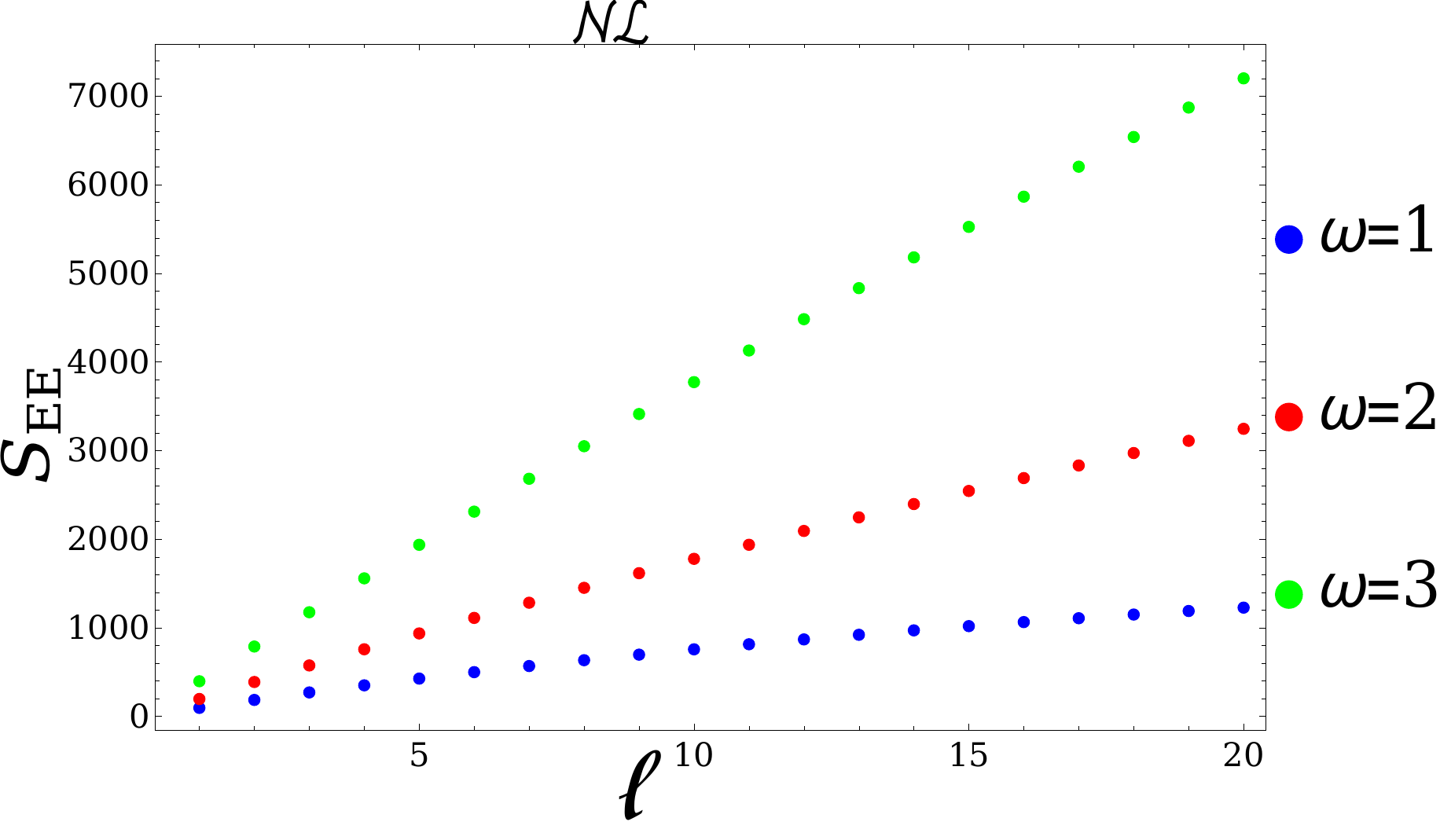}\includegraphics[scale=.28]{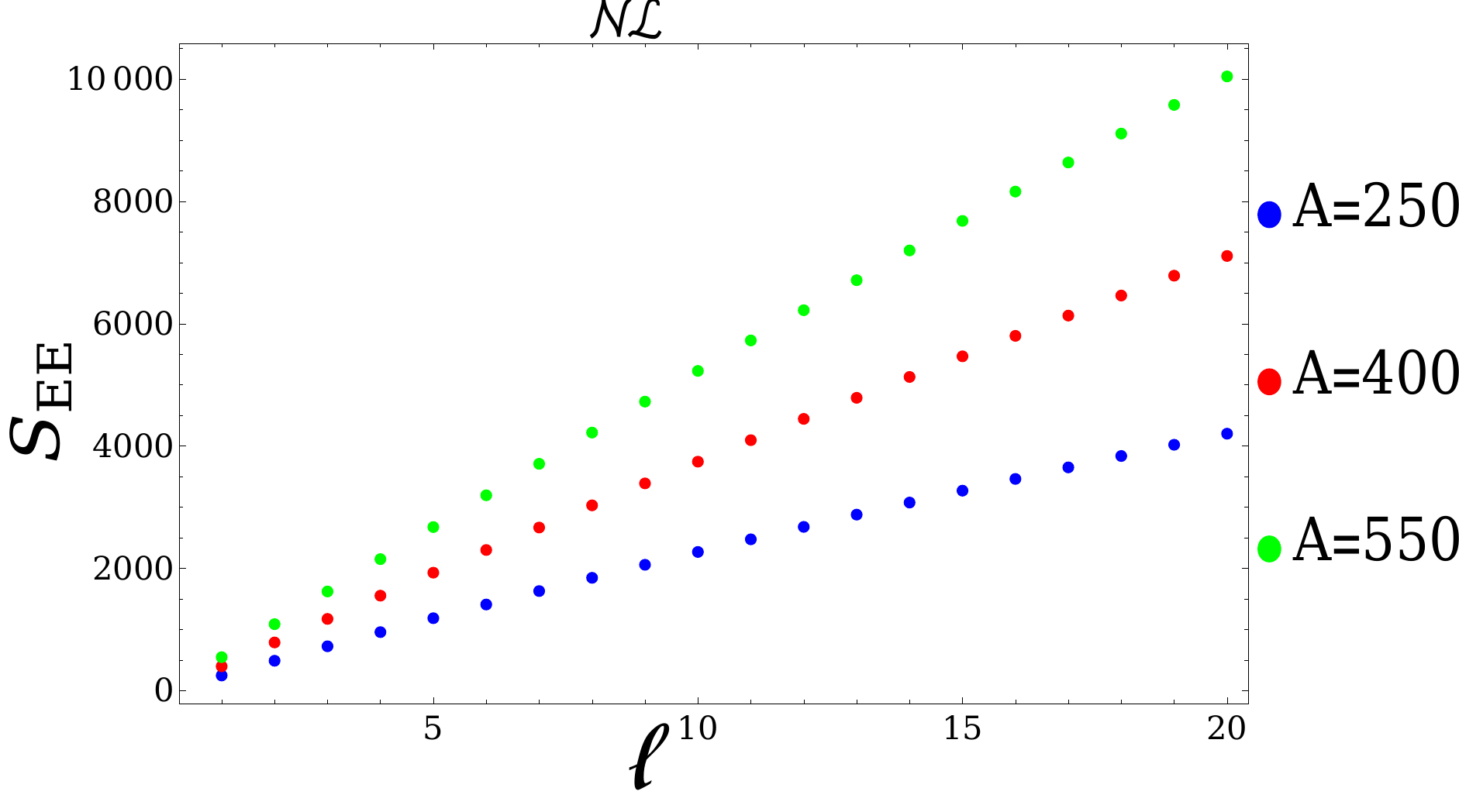} 	 
		\caption{ In the non-local QFT, EE is shown in relation to subsystem length when $ \ell\ll A $. In the left panel, $ A=200$ and  $w=1 $ (blue dots), $w=2$ (red dots), and $w=3$ (green dots) are displayed. On the right panel, $w=2$ and  $A=250 $ (blue dots), $A=400$ (red dots), and $A=550$ (green dots) are shown. It can be observed that the EE obeys the volume law within this range. }  \label{Fig:Seevol}
	\end{figure}

\begin{figure}[h]
		\centering
		\includegraphics[scale=.31]{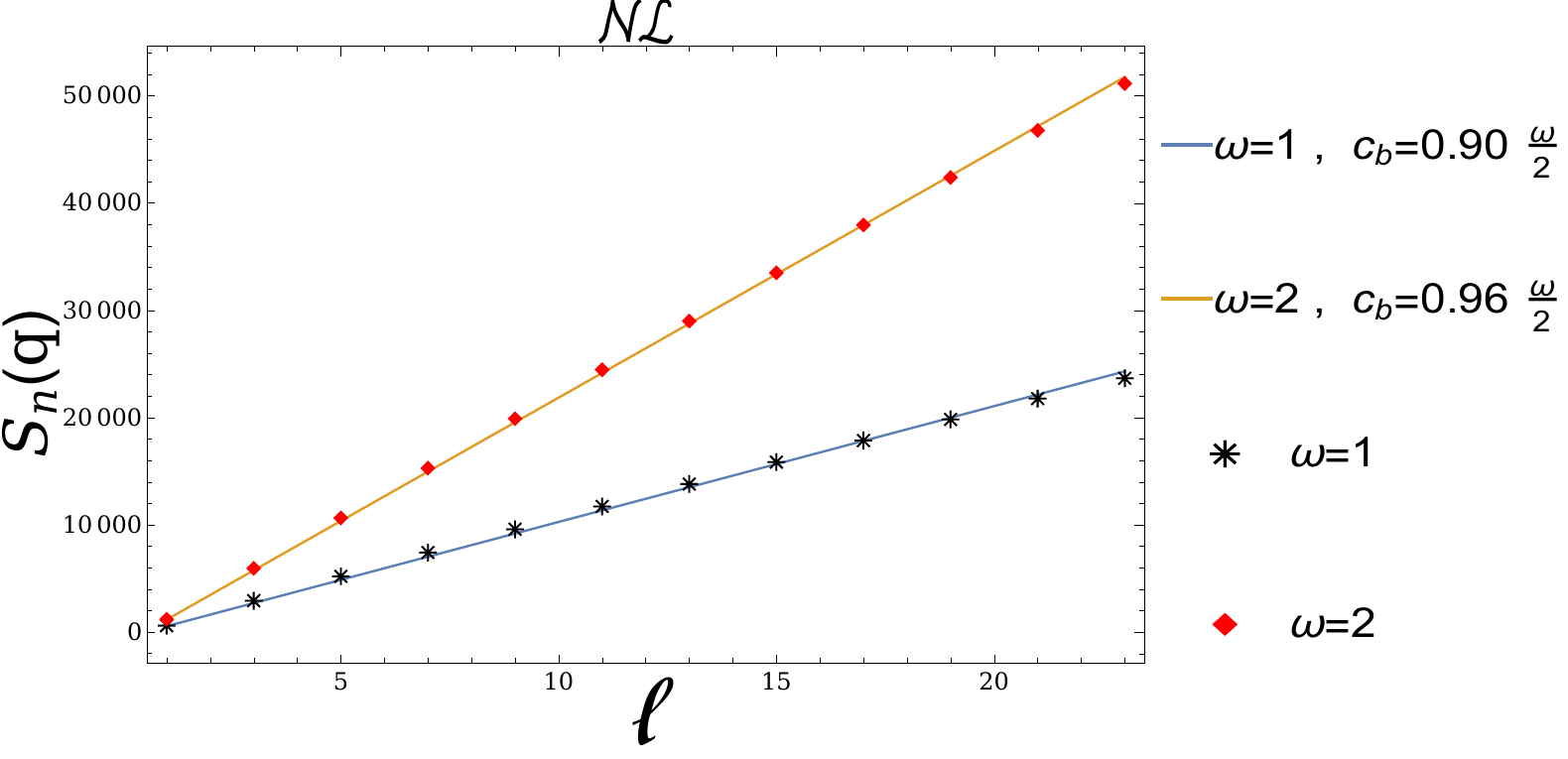} \includegraphics[scale=.31]{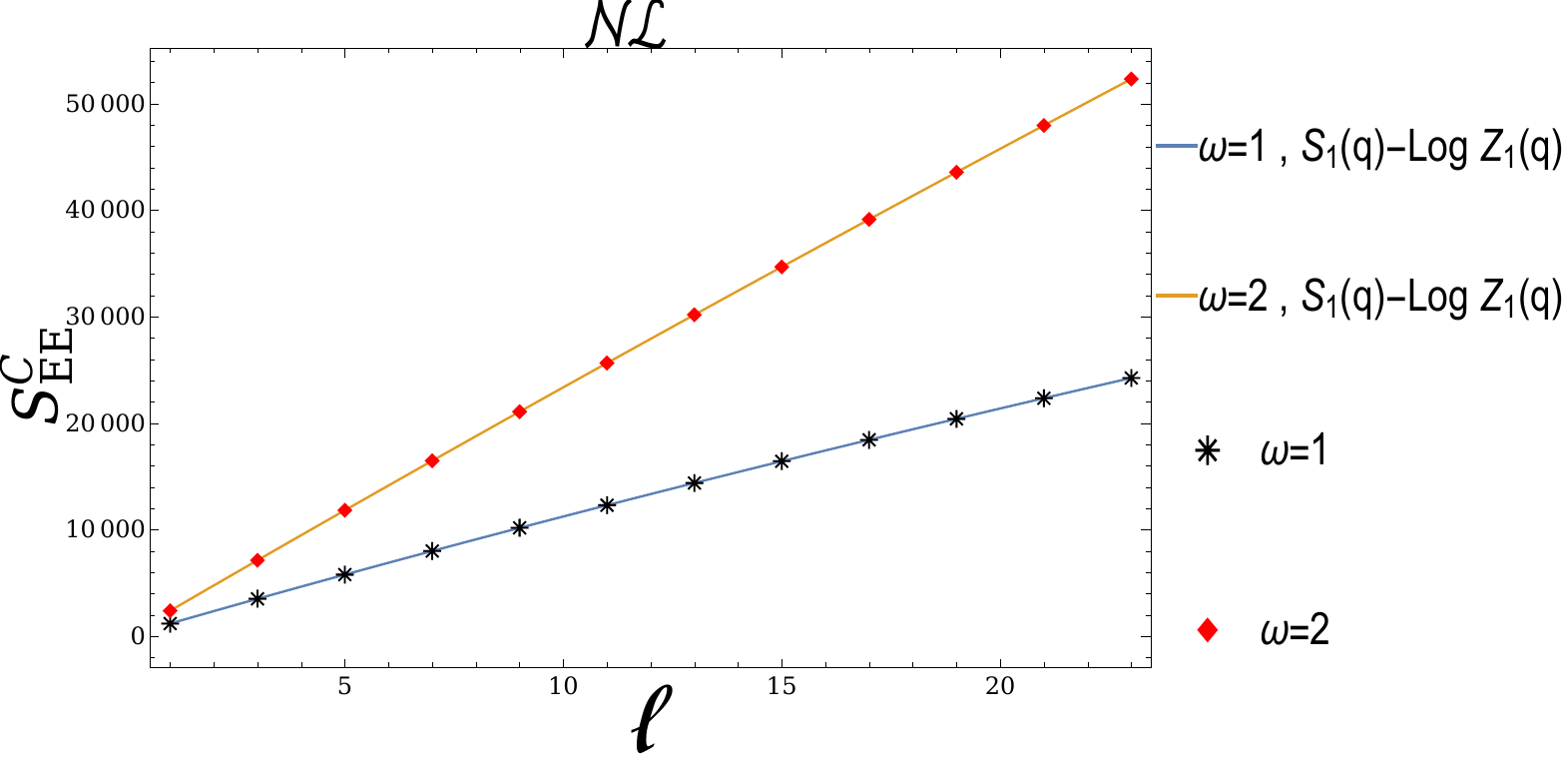} 	 
		\caption{ The EE and symmetry resolved entropies for subsystem size $\ell$ are displayed for $w = 1$ and $2$ in the case of $A=1200$. The solid lines representing the predicted results closely align with the numerical values indicated by the symbols, showing good agreement with values very close to $c_b = \frac{w}{2}$.} \label{Fig:SNQNLana}
	\end{figure}

	To compute the EE, it is necessary to utilize the relationships \eqref{Eq:weqal1}, \eqref{Eq:weqal2}, and $\mathbf{X}_{\mathcal{NL}}$ in place of \eqref{Eq:towpointPKG} and \eqref{Eq:towpointQKG}. To achieve this, we first present Figure \ref{Fig:SeeaNL}, which illustrates the EE for the non-local QFT in the regime where $ \ell \gg A $. Our results successfully reproduce the findings of Ref. \cite{Shiba:2013jja}. Specifically, the right panel of Figure \ref{Fig:SeeaNL} demonstrates that the scaling of $S_{EE}$ follows $A^2$. For small values of $ \ell $ compared to $ A $, the graph in Figure \ref{Fig:Seevol} illustrates the relationship between $S_{EE}$ and $ \ell $, with the scaling given by $S_{EE} = c_b \ell A $, where $c_b$ is a constant dependent on $w$. When $A$ is sufficiently large, a good approximation for $c_b$ is $\frac{ w}{2}$. Furthermore, as the parameters $A$ and $w$ are increased, the value of $S_{EE}$ increases significantly, and its behavior follows the volume law. \\	
To conclude this section, we provide additional comments on determining the $c_b$ coefficient, as well as the configurational and number entropies introduced in \eqref{Eq:SvN}. Using equation \eqref{Eq:SvN}, we express the EE as follows:
\begin{equation}
	\label{Eq:SeeAlarge}
	S^{C~{A\gg \ell}}_{EE}= S^{A\gg \ell}(q)-\log\mathcal{Z}^{A\gg \ell}_1(q),
\end{equation}
By substituting equations \eqref{eq:ZnqAlarge} and \eqref{Eq:SnqAlarge1} into the above equation, we obtain $S^{C~ {A\gg \ell}}_{EE}\simeq \ell \,w\, A$, thereby fixing $c_b\sim \frac{w}{2}$. Numerical verification is presented in Figure \ref{Fig:SNQNLana}, where the right panel shows that the predicted results for $S^{C~{A\gg \ell}}_{EE}$ in equation \eqref{Eq:SeeAlarge} (solid lines) align with the numerical results (symbols) obtained via equation \eqref{Eq:SEE}.

Finally, using equations \eqref{Eq:SvN} and \eqref{Eq:SeeAlarge}, we can argue that equation \eqref{Eq:SnqAlarge1} provides the configurational entropy, while equation \eqref{eq:ZnqAlarge} yields the number entropy, which is approximately $\frac{Aw}{2}$.


	\section{Conclusions}\label{Sec-6}
In this paper, we examined the symmetry-resolved  R\'enyi entropies and SREE for free complex scalar fields and a specific class of non-local quantum field theories whose entanglement entropy satisfies a volume law.  We concentrated on the behavior of the eigenvalues of the correlation matrix for Klein-Gordon fields in both massive and massless limits. By identifying and specifying the most significant eigenvalues of the correlation matrix, we derived effective expressions for the charged moments. Consequently, we demonstrated that the obtained symmetry-resolved entropies are in good agreement with numerical calculations in all cases. Utilizing the estimated eigenvalues, we derived the EE for complex scalar fields in the massless limit, which aligns with CFT results. Additionally, we examined the behavior of symmetry-resolved entropies in relation to both mass and the size of the subsystem, $\ell$. It was demonstrated that $S_n(q)$ saturates for larger masses at smaller values of $\ell$. For massive cases, we verified the $q$-dependence of $\mathcal{Z}_n(q)$ and $S_n(q)$. In contrast, in the massless limit and for large $q$, the dependence of $S_n(q)$ on $q$ becomes more relaxed. Specifically, in the massless limit, $S_n(q)$ are approximately independent of the symmetry sectors, $q$, effectively confirming the equipartition of EE.  We would like to emphasize that our results are consistent with the existing literature. For instance, it has been argued that conformal invariance implies the equipartition of EE in systems with $U(1)$ symmetry \cite{german-3}. Additionally, for generic gapped systems, Ref. \cite{Murciano:2019wdl} reported an effective equipartition of EE. \\
Furthermore, we investigated the SREE in a specific class of non-local QFTs, where the non-locality is characterized by two parameters, $A$ and $w$. In Ref. \cite{Shiba:2013jja}, it was argued that for $ \ell \ll A $, the EE follows $S_{EE} = c_b \ell A$, while for $\ell \gg A$, the EE is proportional to $A^2$. We examined the EE for these theories and presented an effective expression for it. Our observations indicate that for larger values of $w$ and $A$, the coefficient $c_b$ approximates to $\frac{w}{2}$. In the case of SREE for $ \ell \ll A $, we derived effective expressions for the symmetry-resolved partition function $\mathcal{Z}_n(q)$. Consequently, we obtained the symmetry-resolved R\'enyi entropies $S_n(q)$ and, upon comparison with numerical calculations, demonstrated that $S_n(q)$ are independent of both $n$ and $q$. This confirms that the considered class of non-local QFTs, when the size of the subsystem is smaller than $A$, adheres to the equipartition of EE.

For future research, within the class of non-local QFTs, we aim to compute additional measures of entanglement, such as mutual information and tripartite information, and to investigate the well-known inequalities between them. Furthermore, significant insights could be gained from exploring the complexity of this class of non-local QFTs.

	\subsection*{Acknowledgments}
We wish to thank M. Alishahiha, A. Naseh, A. Mollabashi, M. Reza Mozaffar and B. Taghavi for their fruitful comments. We would like to acknowledge M. Ghasemi and G. Jafari for their cooperation in the first part at earlier versions of this paper, hereby,  we appreciate their efforts. We thank the IPM-Grid computing group for providing computing and storage facilities.  R. Pirmoradian would also like to extend special thanks to M. Reza Lahooti Eshkevari dean of Ershad Institute and M. Saidi for their support.

	\appendix

	\section{Comments on a Single Site } \label {single site}
	
\begin{figure}[h]
		\centering
		\includegraphics[scale=.28]{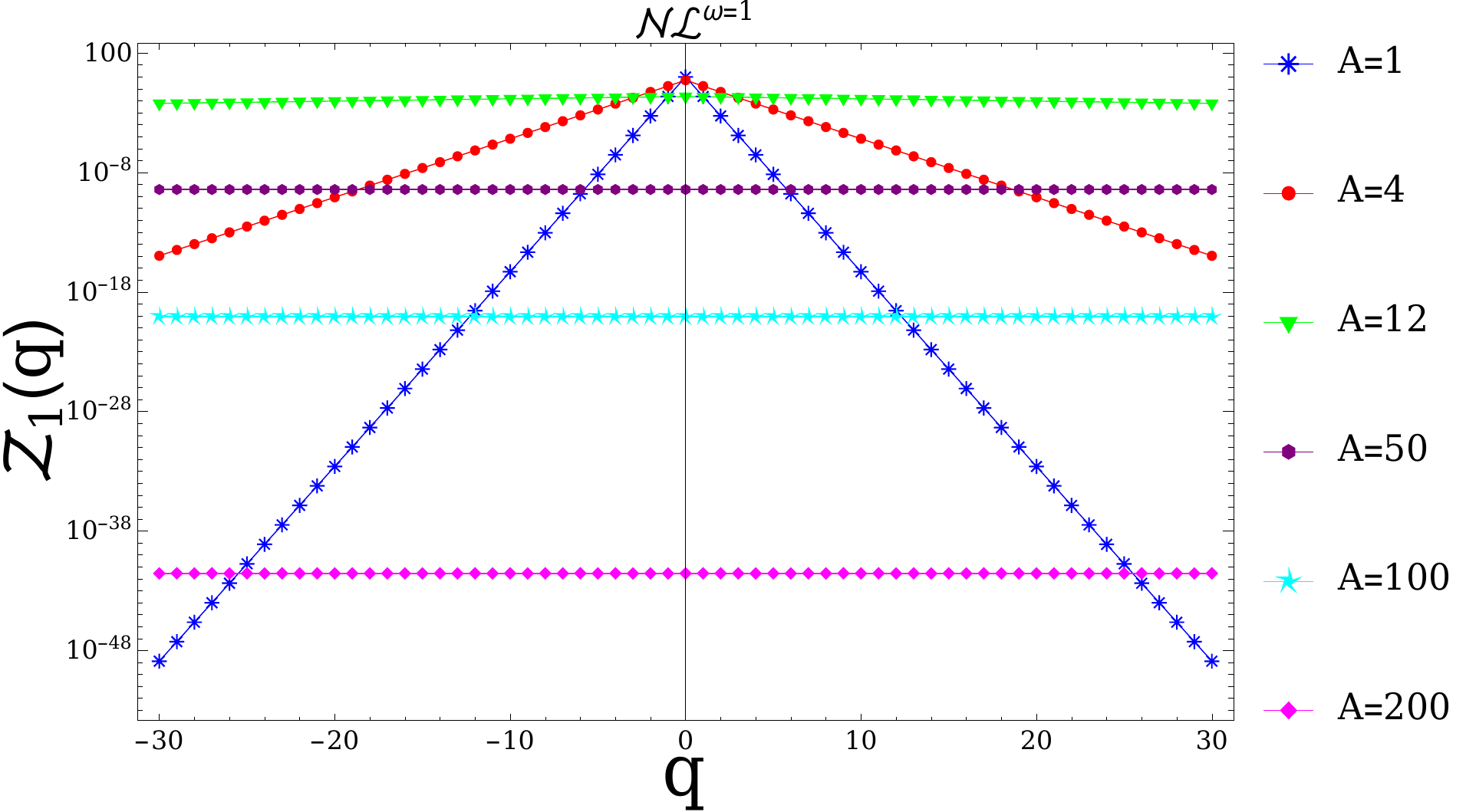}\includegraphics[scale=.28]{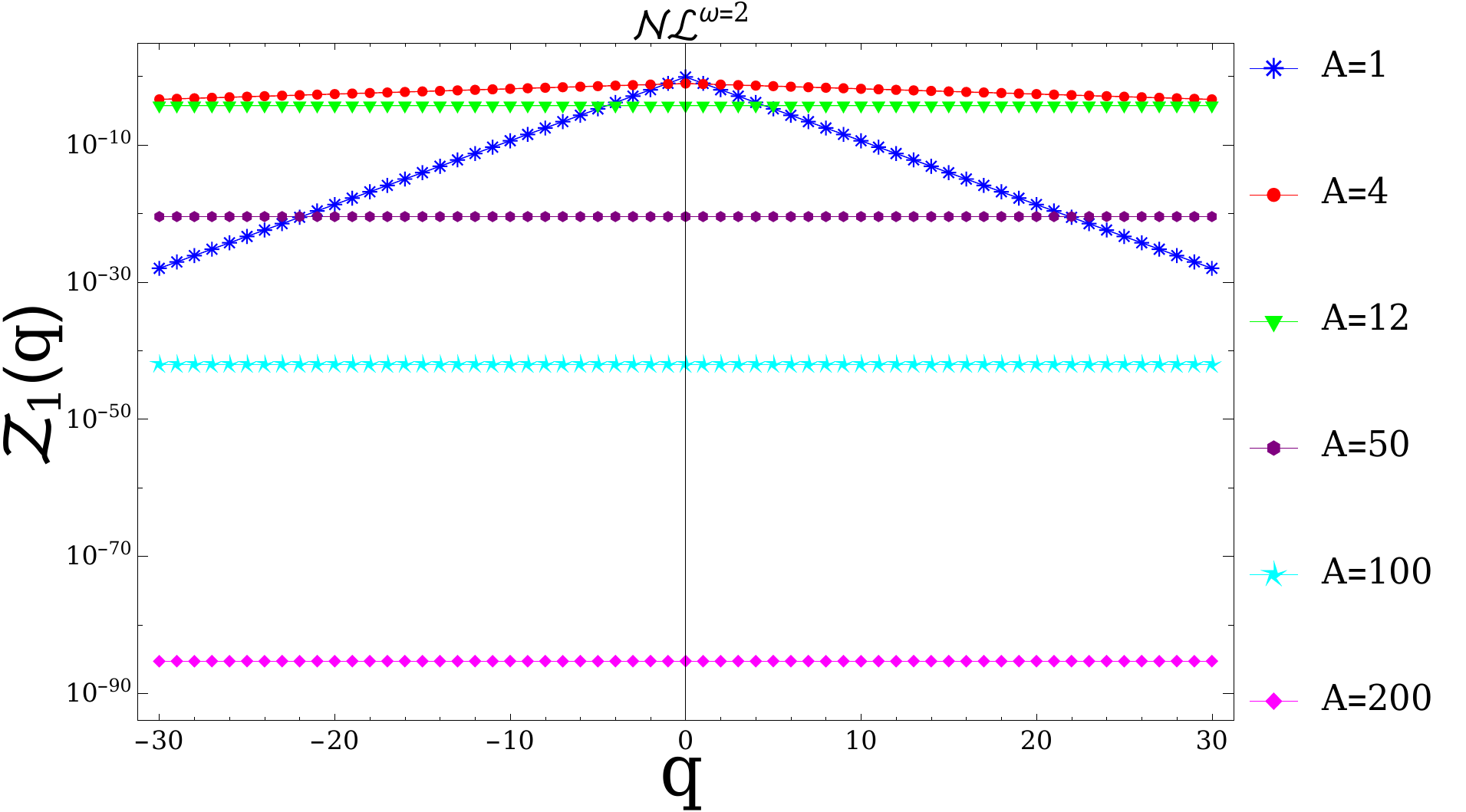} 	 
		\caption{ The single-site graphs illustrate the values of $\mathcal{Z}_1(q)^{\mathcal{NL}}$ as a function of $q$ for $A = 1, 4, 12, 50, 100,$ and $200$ at $w = 1$ (left plot) and $w = 2$ (right plot). It is evident from the figures that the slope of the line decreases as the non-locality parameters ($A$ and $w$) increase. When $A \gg 1$, the value of $\mathcal{Z}_1(q)^{\mathcal{NL}}$ becomes independent of $q$.		
			 } \label{Fig:zeroNL}
	\end{figure}

In this appendix, we provide an expression for the eigenvalues of the correlator matrix for a single site of bosonic scalar QFTs. For a single degree of freedom on a finite lattice and large $\mathcal{N}$, the eigenvalues are obtained as follows:
	\bea
	\nu=\frac{1}{\mathcal{N}}\sqrt{1+\frac{\mathcal{R}_2}{m}+m\mathcal{R}_1+\mathcal{R}_1\mathcal{R}_2}
	\eea 
	where $\mathcal{R}_1$  and $\mathcal{R}_2$ are defined by 
	\bea
	\mathcal{R}_1=\displaystyle\sum_{k=1}^{\mathcal{N}-1}\frac{1}{\omega_k}\,,\hspace{4mm} \hspace{4mm}
	\mathcal{R}_2=\displaystyle\sum_{k=1}^{\mathcal{N}-1}\omega_k.
	\eea
	The massless and the massive limits can be obtained where in the case of the massless one obtains 
	\begin{equation}
		\hspace{4mm} \mathcal{R}_1\mathcal{R}_2\sim4\mathcal{N}^2\,,
		\hspace{4mm}\mathcal{R}_2\sim \mathcal{N}\,,\hspace{4mm}m\mathcal{R}_1\sim 0 
	\end{equation} 
Therefore, the eigenvalues take the following form
	\begin{equation}                         
		\nu _{ massless}=
		\frac{1}{\mathcal{N}} \sqrt{1+\frac{\mathcal{N}}{m}+4\mathcal{N}^2}\sim\frac{1}{\sqrt{m\mathcal{N}}}.
	\end{equation}
For the  massive limit, one obtains
	\begin{equation}
		\frac{\mathcal{R}_2}{m}=m\mathcal{R}_1\sim\mathcal{N},\hspace{4mm} \mathcal{R}_1\mathcal{R}_2=\mathcal{N}^2  ,
	\end{equation}  
leading to the following eigenvalue 	
	\begin{equation}                         
		\nu _{ massive}=
		\frac{1}{\mathcal{N}} \sqrt{1+2\mathcal{N}+\mathcal{N}^2}\sim1+\frac{1}{\mathcal{N}}
	\end{equation}
The calculations can be extended to the case where $\mathcal{N}$ approaches infinity, yielding the following result:
	\begin{equation}                        
		\nu _{ massless}=
		\frac{2}{\pi} \sqrt{\log \left(\frac{8}{m}\right)}+O\left(m^2\right)
	\end{equation}
where leads to the following symmetry-resolved partition function
	\begin{equation}
		\mathcal{Z}_1(q)=
		\frac{\pi}{2}\frac{1}{ \sqrt{\log \left(\frac{8}{m}\right)}}  \left(1-\frac{2 \pi }{2 \sqrt{\log \left(\frac{8}{m}\right)}+\pi }\right)^{\left| q\right| }
	\end{equation}
\begin{figure}[h]
		\centering
		\includegraphics[scale=.28]{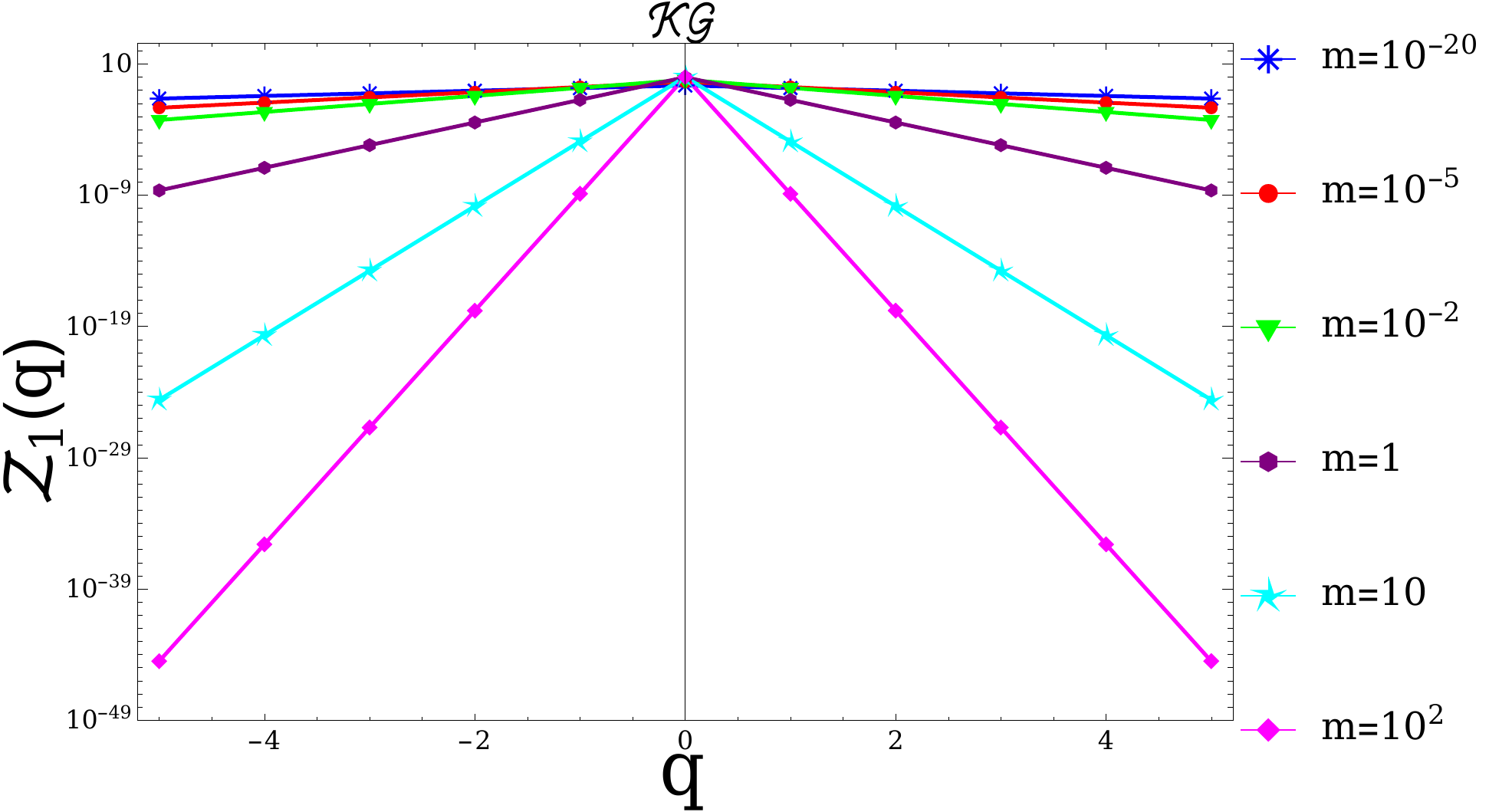}\includegraphics[scale=.28]{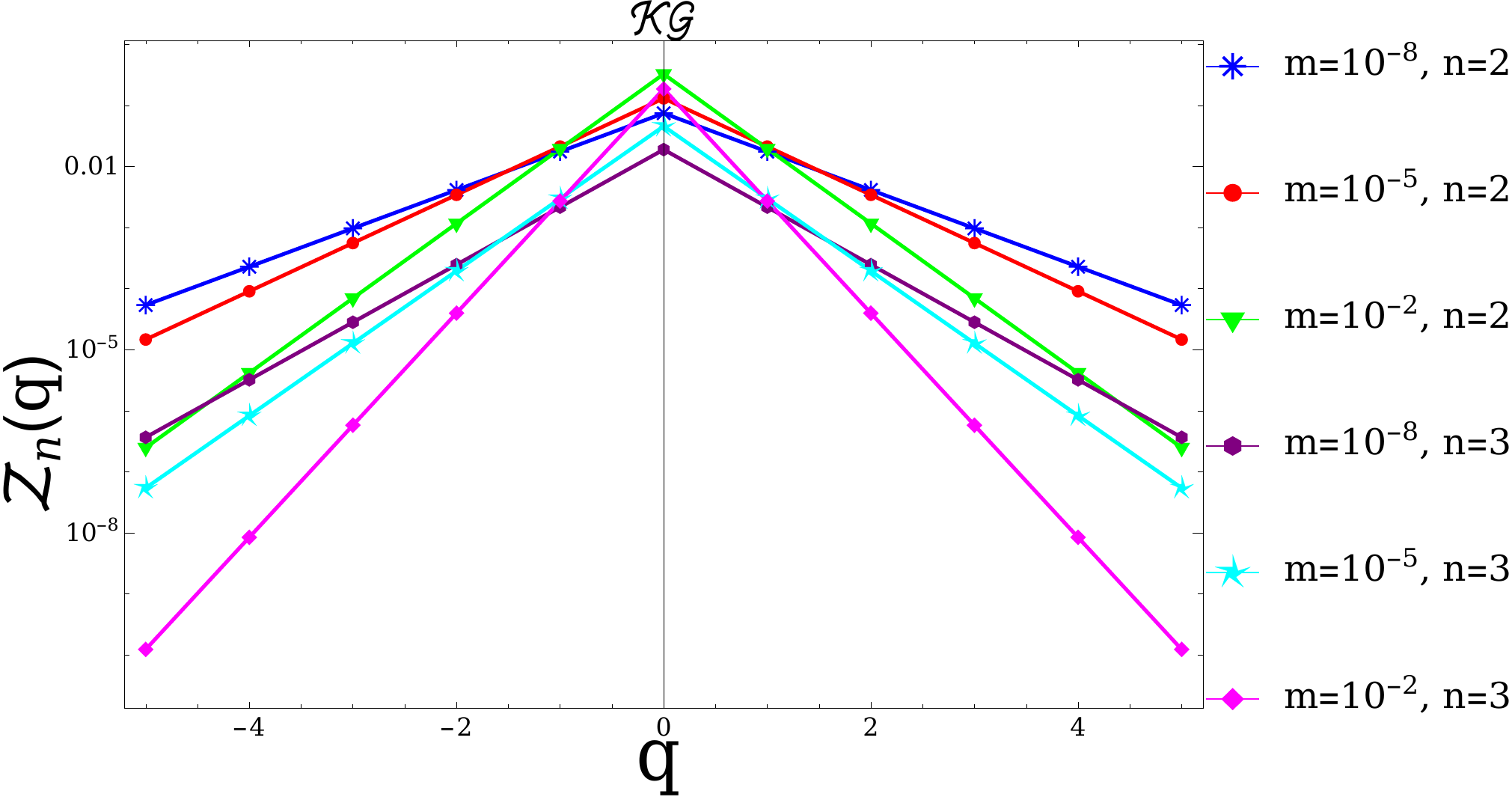} 	 
		\caption{The values of $\mathcal{Z}_n(q)$ for a single-site, plotted against $q$, are presented for $m = 10^{-20}, 10^{-5}, 10^{-2}, 1, 10,$ and $10^{2}$ for $n = 1$ (left plot) and for $m = 10^{-8}, 10^{-5}, 10^{-2}$ for $n = 2, 3$ (right plot). It is evident that the slope of the line decreases as the mass decreases. Consequently, for $m \ll 1$, the value of $\mathcal{Z}_1(q)$ becomes independent of $q$.}  \label{Fig:zero}
	\end{figure}
And in the massive limit, one obtains  
	\begin{equation}
		\nu _{ massive}=
		1+\frac{1}{4 m^4}+O\left(\frac{1}{m}\right)^6
	\end{equation}
thus $	\mathcal{Z}_1(q)$ is obtained as 	
	\begin{equation}
		\mathcal{Z}_1(q)=
		\frac{4 m^4 \left(8 m^4+1\right)^{-\left| q\right| }}{4 m^4+1}\approx\left(8 m^4+1\right)^{-\left| q\right| }
	\end{equation}
For the non-local quantum fields at $\mathcal{N}\to\infty$ and for $ w=1  $, one obtains	
	\begin{equation}
		\nu^{\mathcal{NL}\,,w=1}=
		\sqrt{I_0(A){}^2-\pmb{L}_0(A){}^2}
	\end{equation} 
where $\pmb{L}_0$ is modified Struve function, in this case one has	
	\begin{equation}
		\mathcal{Z}_1(q)^{\mathcal{NL}\,,w=1}=
		\frac{1}{\sqrt{I_0(A){}^2-\pmb{L}_0(A){}^2}}\left(\frac{\sqrt{I_0(A){}^2-\pmb{L}_0(A){}^2}-1}{\sqrt{I_0(A){}^2-\pmb{L}_0(A){}^2}+1}\right){}^{\left| q\right| }
	\end{equation}
	And for $ w=2  $, one obtains	
	\begin{equation}
		\nu^{\mathcal{NL}\,,w=2}=
		I_0(A)
	\end{equation} 
leading to 	
	\begin{equation}
			\mathcal{Z}_1(q)^{\mathcal{NL}\,,w=2}=
		\frac{1}{I_0(A)}\left(1-\frac{2}{I_0(A)+1}\right){}^{\left| q\right| }.
	\end{equation}
The Figures \ref{Fig:zero} and \ref{Fig:zeroNL} present the results for large values of $A$ and $w = 1, 2$, illustrating $\mathcal{Z}_1(q)$ plotted against $q$. The single-site graphs further confirm that the symmetry-resolved moment remains $q$-independent for non-local fields and $A \gg \ell$, a characteristic that is approximately preserved in the massless limit of complex scalar fields.   	


\end{document}